\documentclass[5p]{elsarticle}
\usepackage[utf8]{inputenc}

\usepackage{gensymb}
\usepackage{caption}
\usepackage{subcaption}
\usepackage{wasysym}
\usepackage[british]{babel}
\usepackage{multirow}
\usepackage{tabularx}
\usepackage{breakurl}

\usepackage{tikz}
\usetikzlibrary{patterns}

\newcommand{\eref}[1]{Eq.~\ref{#1}}
\newcommand{\fref}[1]{Fig.~\ref{#1}}
\newcommand{\tref}[1]{Table~\ref{#1}}

\newcommand{\sref}[1]{Section~\ref{#1}}

\newcommand\T{\rule{0pt}{2.6ex}}
\newcommand\B{\rule[-1.2ex]{0pt}{0pt}}

\begin{document}

\begin{frontmatter}
\title{Evaluation of commercial nickel-phosphorus coating for ultracold neutron guides using a pinhole bottling method}

\author[lanl]{R.W.~Pattie~Jr\corref{cor1}}
\ead{rwpattie@lanl.gov}
\author[iu]{E.R.~Adamek}
\author[ill]{T.~Brenner}
\author[ncsu,tunl]{A.~Brandt}
\author[lanl]{L.J.~Broussard}
\author[iu]{N.B.~Callahan}
\author[lanl]{S.M.~Clayton}
\author[ncsu,tunl]{C.~Cude-Woods}
\author[lanl]{S.A.~Currie}
\author[ill]{P.~Geltenbort}
\author[lanl]{T.M.~Ito}
\ead{ito@lanl.gov}
\author[tum]{T.~Lauer\fnref{movtc}}
\author[iu]{C.Y.~Liu}
\author[lanl]{J.~Majewski}
\author[lanl]{M.~Makela} 
\author[hear]{Y.~Masuda}
\author[lanl]{C.L.~Morris}
\author[lanl]{J.C.~Ramsey}
\author[lanl,iu]{D.~Salvat}
\author[lanl]{A.~Saunders}
\author[tum]{J.~Schroffenegger}
\author[lanl]{Z.~Tang}
\author[lanl]{W.~Wei}
\author[lanl]{Z.~Wang}
\author[lanl]{E.~Watkins}
\author[ncsu,tunl]{A.R.~Young}
\author[lanl,ncsu,tunl]{B.A.~Zeck}

\address[lanl]{Los Alamos National Laboratory, Los Alamos, NM 87544, USA}
\address[iu]{Physics Department, Indiana University, Bloomington, IN 47405, USA }
\address[ill]{Institut Laue-Langevin, 6, Rue Jules Horowitz, 38042 Grenoble CEDEX, France}
\address[ncsu]{Department of Physics, North Carolina State University, Raleigh, NC 27606, USA}
\address[tunl]{Triangle Universities Nuclear Laboratory (TUNL), Durham, NC 27708, USA} 

\address[hear]{High Energy Accelerator Research Organization, 1-1 Oho, Tsukuba, Ibaraki 305-0801, Japan}
\address[tum]{Technical University of Munich, Munich, Germany}		
 
\cortext[cor1]{Corresponding author}
\fntext[movtc]{Present Address: Movatec GmbH, t.lauer@movatec.de}

\begin{abstract}
  We report on the evaluation of commercial electroless nickel phosphorus (NiP) coatings for ultracold neutron (UCN) transport and storage. The material potential of 50~$\mu$m thick NiP coatings on stainless steel and aluminum substrates was measured to be $V_F = 213(5.2)$~neV using the time-of-flight spectrometer ASTERIX at the Lujan Center.  The loss per bounce probability was measured in pinhole bottling experiments carried out at ultracold neutron sources at Los Alamos Neutron Science Center and the Institut Laue-Langevin.  For these tests a new guide coupling design was used to minimize gaps between the guide sections. The observed UCN loss in the bottle was interpreted in terms of an energy independent effective loss per bounce, which is the appropriate model when gaps in the system and upscattering are the dominate loss mechanisms, yielding a loss per bounce of $1.3(1) \times 10^{-4}$.   We also present a detailed discussion of the pinhole bottling methodology and an energy dependent analysis of the experimental results.
  
\end{abstract}

\end{frontmatter}

\section{Introduction}

Ultracold neutrons have been recognized as an excellent tool for probing our understanding of the fundamental laws of nature, and new programs of research are fast-developing worldwide. Experiments are planned or have been carried out at UCN sources to measure the neutron's lifetime \cite{Salvat2014a,Serebrov2005d,Pichlmaier2010a,Arzumanov2015,Materne2009,Leung2016} and electric dipole moment \cite{Baker2014,Afach2015,Ito2013,Lamoreaux2009a}, angular correlations in neutron $\beta$-decay \cite{Mendenhall2013,Broussard2016}, and gravitational energy states \cite{Nesvizhevsky2003,Jenke2011a}.  Many of these experiments are statistics limited and several new UCN sources \cite{Martin2013,Anghel2009,Karch2014} in development promise to deliver the required densities to achieve higher precision measurements. To deliver the required statistics for current and future experiments the spallation UCN facility at the Los Alamos Neutron Science Center (LANSCE) \cite{Saunders2013} is upgrading several key components of their source \citep{Ito2013}, including the UCN guide system out of the source.  The primary motivation for this upgrade is to achieve a UCN density $>10$ UCN/cc, which will enable a future room temperature neutron electric dipole moment search with a goal sensitivity of $3 \times 10^{-27}$~e$\cdot$cm.

UCN have sufficiently low kinetic energy can be totally internally reflected in material bottles. The interaction of the UCN and a material wall is described by the effective Fermi potential
\begin{equation}
	V_{F} = V - i W = \frac{2\pi \hbar^2}{m_n} N a - i \frac{\hbar}{2}N \sigma v,
	\label{eq:vfermi}	
\end{equation} 
where $m_n$ and $v$ are the mass and velocity of the neutron, $N$ is the number density of the atoms in the material, $a$ is the coherent neutron scattering length of the material, and $\sigma$ is the loss cross section (See e.g. Ref.\cite{UCNBook} and the references found therein). The real part of the Fermi potential gives the maximum energy of neutrons that can be confined, and the imaginary part describes the absorption and upscattering. The ratio of the imaginary and real parts of $V_F$, $f = W/V$, is defined as the loss factor.  The loss probability $\mu(E)$, as a function of the UCN kinetic energy $E$, for a bottle of material potential $V$ is given by 
\begin{equation}
 \mu(E) = 2 f \left[  \frac{V}{E}\sin^{-1}\left( \frac{E}{V} \right)^{1/2} - \left( \frac{V}{E} -1 \right )^{1/2} \right],
 \label{eq:loss}
\end{equation}
after averaging over all angles of incidence \citep{Golub1979}. Typically, the energy scale of $V_F$ is 100's of neV with $^{58}$Ni possessing the largest Fermi potential $V_{F} = 342$~neV among commonly available materials.  A convenient quirk of nature sets the energy scale for interaction with magnetic fields, $V_{mag}=- \vec{\mu} \cdot \vec{B} \sim 60$~neV/T, and earth's gravity, $V_{g} = m_n g h \sim 100$~neV/m, on similar footing.  Experiments have exploited these interactions to fully polarize a ``beam" of UCN using a large magnetic field or significantly alter the energy of UCN by raising or lowering their apparatus a meter or two relative to the beam height.

In the LANSCE source, 6~m of electro-polished stainless steel tubes guide the UCN from the source volume to the experimental area.  Stainless steel has an effective material potential of $V_F=188 $~neV. The current guide system has a loss-per-bounce probability of $5.2\times10^{-4}$ \cite{Saunders2013}, which includes the effect of gaps between the 1~m guide sections.  Improving the transport from the source volume to the experimental hall will have a significant impact on achieving desired UCN densities.
   
Pure nickel coatings provide a  high Fermi potential for bottling UCN; however, the probability of spin-flip is unacceptably high for applications requiring polarized UCN.  Nickel alloy coatings, such as nickelmolybdenum
or nickel-vanadium, have been shown to have high Fermi potential (210-220 neV), high secularity, and low spin-flip probability \citep{Daum2014a}.  Electroless nickel plating has been used since the 1950's to provide robust corrosion resistant uniform coatings on conductive and non-conductive substrates.  Unlike electro-plating, this method uniformly coats objects with complex geometries.  When the phosphorus content is in excesses of 10\% by atomic composition the coating is non-magnetic at room temperature\cite{Humbert1998,BERRADA1978,Albert1967}.  The combination of the robust nature of the coating, the commericial availability, high Fermi potential, and the non-magnetic surfaces make nickel phosphorus (NiP) an ideal candidate for a UCN transport guide coating material.  NiP coated guides have been in use at the UCN source in Osaka for more than a decade \cite{Masuda2002}. Our group recently reported on a measurement of the spin-flip probability for UCN interacting with NiP coated stainless steel and aluminum neutron guides \cite{Tang2015}.  The results of that study show the spin-flip probability to be $\beta = 3.3^{1.8}_{-5.6}\times 10^{-6}$, which is sufficient for experiments requiring the transport and storage of polarized UCN.

In this paper we report on a suite of measurements of the bottle lifetime of NiP coated stainless steel guides performed at the PF2 facility at the Institut Laue Langevin \cite{PF2ILL} and LANSCE UCN \cite{Saunders2013} sources.  The purpose of the measurements was to evaluate the overall performance of the guide technology chosen for our new UCN source, including the surface properties for UCN as well as the guide coupling design. In Sec. \ref{sec:vfermi} the results of a measurement of the real part of the Fermi potential and an estimate of the imaginary part are presented.  Section \ref{sec:guide} summarizes the guide preparation and coating. Section \ref{sec:expmeth} details the double pinhole method.  Sections \ref{sec:proc} and \ref{sec:ILL} present the measurements performed at LANL and the ILL, respectively.  Section \ref{sec:disc} provides a discussion of our results.  

\section{Determination of the NiP Fermi Potential}
\label{sec:vfermi}
\subsection{Measurement of the Real Part of $V_F$}

The Fermi potential of NiP coated surfaces was measured using the time-of-flight spectrometer ASTERIX at the Lujan Neutron Scattering Center at LANSCE. ASTERIX views an intense polychromatic neutron beam through a neutron guide with a 36~cm$^2$ cross-sectional area and utilizes the wavelength band $\lambda$ from 4 to 13 \AA. The divergence of the neutron beam and its footprint on the beam at the sample can be controlled with two sets of slits. Two aluminum and two stainless steel coupons, with surface area of 50$\times$50~mm$^2$, were coated with 50$\mu$m thick films of nickel phosphorus by Chem Processing, INC \cite{chemprocessing}  using the electroless process described in \sref{sec:guide}.  Neutron reflectometry experiments were performed at the neutron incidence beam angle $\theta$ of 0.7$^{\circ}$. To minimize the effect of surface roughness, mostly from polishing and machining,\footnote{Roughness here is defined as combined effect of sample surface deviation from mathematically sharp interface.} the beam was aligned along the polishing marks.

\begin{figure}
\centering
\includegraphics[width=0.45\textwidth,viewport=10 0 190 175,clip=true]{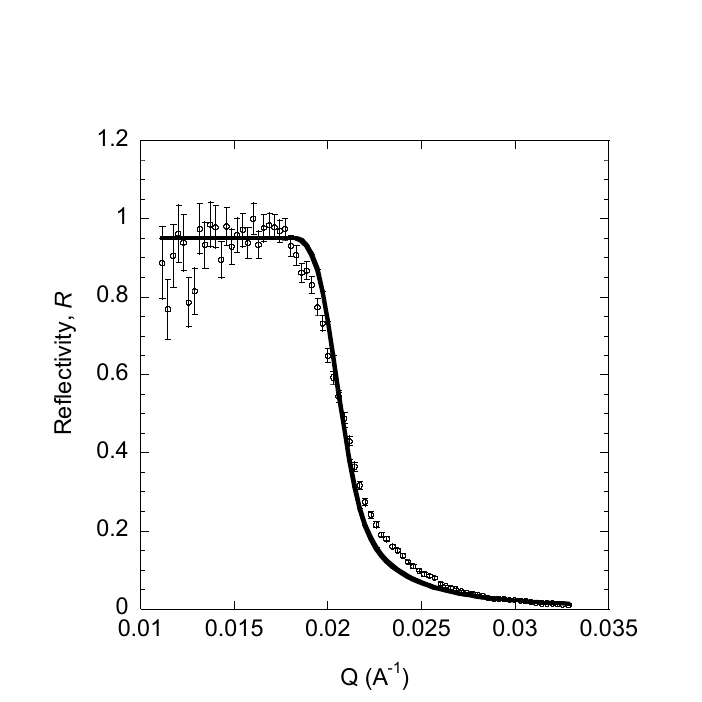}
\caption{\label{fig:asterix} The fitted reflectivity curve (solid line) for the sample 3. The scattering length density of the film was 8.2(2)$\times 10^{-6}$ \AA$^{-2}$. }
\end{figure}	

The obtained  $I(\theta ; \lambda)$ data was reduced and binned 
 to obtain the reflected beam intensity, $R$, as a function of the neutron momentum transfer vector $Q$.  $R$ is normalized to the intensity of the incoming beam (and therefore equal $\approx 1$ below the critical momentum transfer vector, $Q_c$) and $Q= 4 \pi \sin \theta /\lambda$.  \fref{fig:asterix} shows the reduced $R(Q)$ results for the NiP coated on stainless steel sample.

The $R(Q)$ curves were fit using an open-source reflectivity package, MOTOFIT, which runs in the IGOR Pro environment \cite{Nelson2006b}. Using the Abeles matrix formalism, a theoretical reflectometry curve can be calculated and compared to the measured data. Both generic optimization and Levenberg–Marquardt nonlinear least-square methods were employed to obtain the best fits with the lowest $\chi^2$ values. The only two parameters varied in the fitting procedure were the scattering length densities of the thin coating films and the beam normalization parameters to account for imperfect beam normalization.  The deviation of the data from the fit at the inflection points is due to imperfections in the sample and does not affect the determination of the mean scattering length density used to calculate the Fermi Potential.

The values of the scattering length density obtained from the measured critical momentum transfer are shown in \tref{tab:nscat}. The scattering length density $N_b$ is related to the Fermi potential describe in \eref{eq:vfermi} as $N_b = N a$, the product of the number density and the scattering length.  The Fermi potential obtained from the measured scattering length density is also listed in \tref{tab:nscat}. The measured Fermi potential is consistent with our expectation based on the phosphorus content and density of NiP found in the literature.

\begin{table}
 \center
 \caption{Results of scattering length density measurement at the ASTERIX spectrometer. Four samples were tested: two aluminum coupons and two stainless steel coated with 50~$\mu$m of nickel-phosphorus.  $N_b$ is the measured scattering length density and $V_F$ is the Fermi potential.  The uncertainty in the measurement of the scattering length density was dominated by the surface roughness of the samples. \label{tab:nscat}}
 \begin{tabular}{cl|c|c}
 	\hline
 	\hline
 		\T Sample & Description & $N_b \times 10^{-6}$ [\AA$^{-2}$] & $V_F$ [neV] \\
 	\hline
 	 \T	1 & NiP on Al & 8.15(20) & 212.2(5.2) 	\\
 	    2 & NiP on Al & 8.10(20) & 210.9(5.2) 	\\
 	    3 & NiP on SS & 8.20(20) & 213.5(5.2) 	\\
 	 \B 4 & NiP on SS & 8.20(20) & 213.5(5.2) 	\\
 	\hline
 	\hline
 \end{tabular}
\end{table}

\subsection{Calculation of the Nickel Phosphorus loss factor}

The imaginary part of the Fermi potential for the NiP coating can be estimated using \eref{eq:vfermi} and summing over the constituent elements weighted by their density in the mixture,
\begin{equation}
	W = \frac{\hbar}{2} \sum_i N_i \sigma_l^{(i)} v.
\end{equation}
In this calculation the measured thermal neutron absorption cross section, $\sigma_{abs}$, is used for the loss cross-section, neglecting contributions from inelastic scattering ($\sigma_{abs}^{Ni} = 4.49$~b and $\sigma_{abs}^P = 0.172$~b) \cite{NeutronNews1992}.  The phosphorus content of the coating is 10.5(25)\% by weight. The NiP  density is 7.8~g/cm$^3$ \cite{Sudagar2013}.  Using the measured Fermi potential this results in estimates for the imaginary part of the potential of $W=2.30(6) \times 10^{-2}$~neV and a loss factor of $f = 1.08(4) \times 10^{-4}$, where the uncertainty is dominated by the variation in the ratio of phosphorus to nickel in the coating.

\section{Guide Preparation}
\label{sec:guide}
\begin{figure}
	\centering
	\includegraphics[width=0.45\textwidth,clip=true,viewport=100 50 600 550]{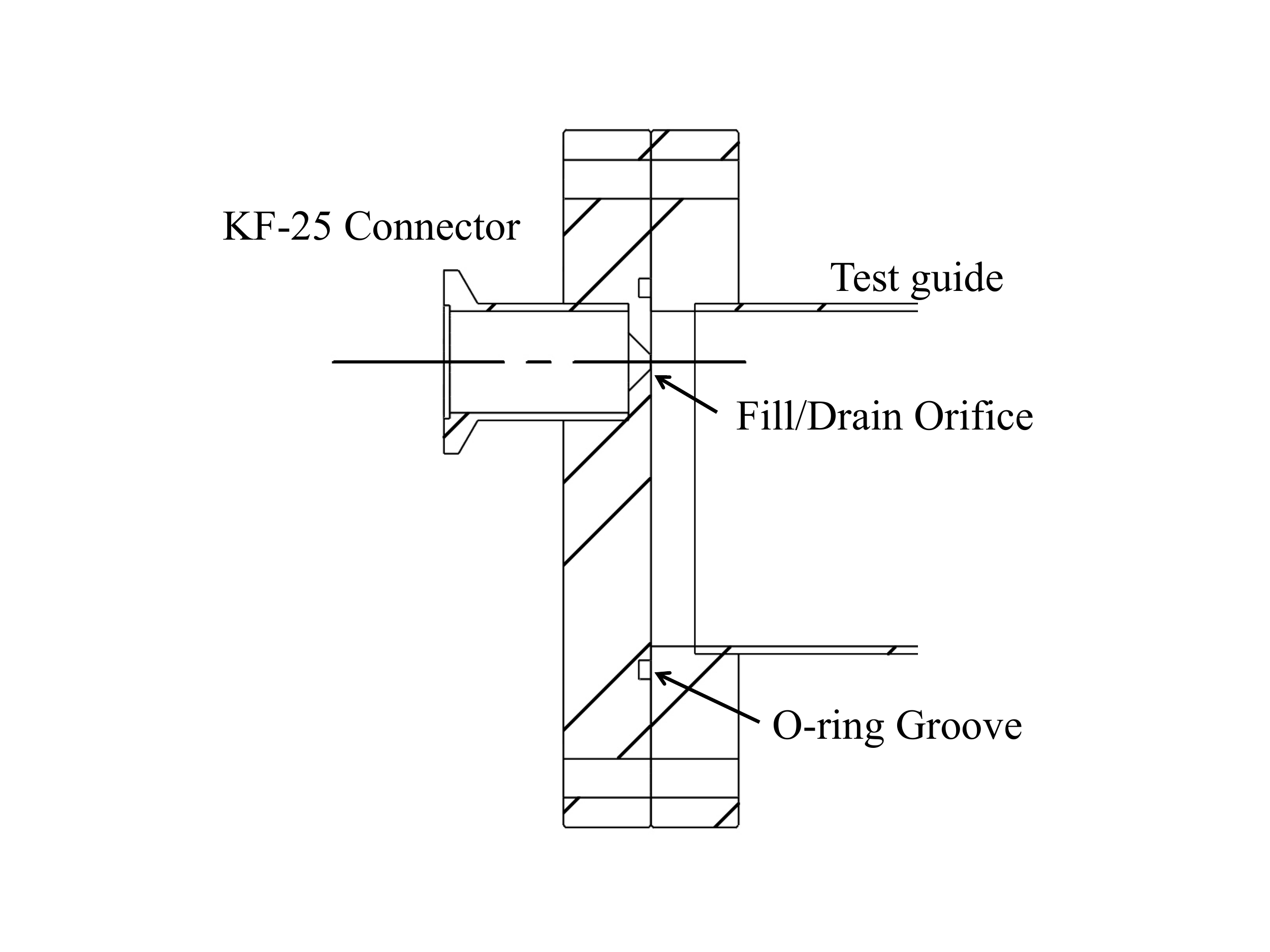}
	\caption{\label{fig:pinhole1} Schematic of the pinhole flange mating to the guide flange. A standard KF-25 nipple, welded to the flange, allows the test guide to be connected to a UCN guide or a $^{10}$B ZnS:Ag detector assembly.}
\end{figure}

\begin{figure}
	\centering
	\includegraphics[width=0.45\textwidth,clip=true,viewport=100 50 600 550]{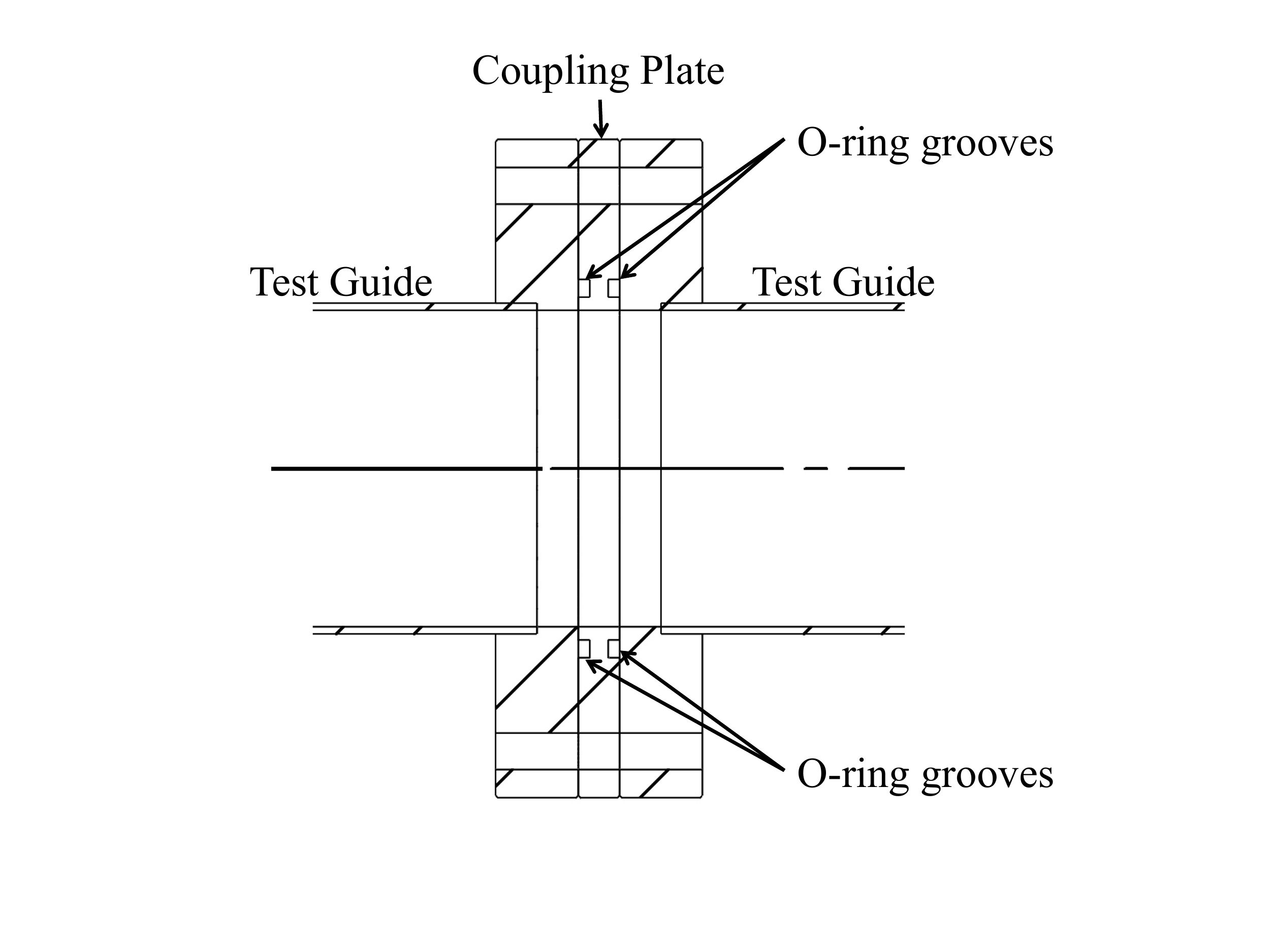}
\caption{\label{fig:coupling}Schematic of test guides mated together using a coupling plate.  Grooves cut in the coupling plate hold o-rings which seal against of the flat surface of the flange.  The coupling plate was coated in NiP.  }
\end{figure}

The UCN guides were constructed from 7.62~cm~OD, 7.29~cm~ID Valex \cite{Valex} stainless steel tubing. Four guides were fabricated, denoted 1-4 in the following.  Guides 1 and 2 were 70 cm long and guides 3 and 4 were 140 cm in length. In the new guide coupling design the tube was welded into a socket, and the surface was smoothed to minimize any surface impact at the weld joint, as shown in \fref{fig:pinhole1}.  In this configuration, guides mate to each other by inserting a thin coupling plate between them, as shown in \fref{fig:coupling}. The flush surface is important to limit gaps, an important source of loss in UCN transmission. After machining, the guides were cleaned with a warm water (63$^{\circ}$ C) Alconox solution (1\% by weight) and rinsed with de-ionized water and isopropyl alcohol. All guides were coated with a high phosphorus\footnote{The phosphorus content in Chem Processing INC's high phosphorus coating is 10.5(25)\% by weight.} NiP mixture by ChemProcessing, INC.   Guide 4 was electro-polished prior to coating \footnote{The Valex tubing used was electro-polished after it was manufactured.  The electro-polishing referred here is after machining to length and welding the mating flanges to the tube.}.  The coating process is an electroless chemical dip that uniformly coats all surfaces of the guide with 50~$\mu$m of NiP, described in more detail in Ref. \cite{Tang2015}.  This process exacerbates the inherent roughness of the substrate which can increase the probably of diffuse scattering. The coupling plate and pinhole flanges were coated in the same NiP as the guides.     

\section{Experimental Method}
\label{sec:expmeth}

\subsection{Double pinhole approach to loss probability measurement}

Focusing on the specific example of this technique described in Brenner et al \cite{Brenner2015}, a measurement of the loss probability was performed by trapping UCN in a bottle made of a test guide and shutters on the ends.  The upstream shutter is opened and the downstream shutter is closed to fill UCN into the bottle.  After the UCN density in the bottle saturates (when the fill rate is balanced by the loss rate), the upstream shutter is closed. Both shutters remain closed for a holding time varied from a few seconds to several hundred seconds and the surviving UCN are emptied and counted by opening the downstream shutter.  The lifetime of the bottle is then determined from a double exponential fit to the number of unloaded UCN versus holding time, where the short $\tau_s$ and long $\tau_l$ time constants are attributed to the lifetimes of untrappable and trappable UCN.  Untrappable UCN are assumed to have kinetic energy greater than the Fermi potential of the bottle walls and the loss rate of this class of UCN is dominated by the time required to have a near-normal incidence collision with the walls, which is influenced by the probability of diffuse scattering.  

The measured long time constant $\tau_l$ is related to the bottle lifetime $\tau_{bottle}$ by
\begin{equation}
	\tau_l^{-1} = \tau_\beta^{-1} + \tau_{bottle}^{-1} + \tau_{gap}^{-1} + \tau_{shutter}^{-1},
	\label{eq:dblegvloss}
\end{equation}
where $\tau_{bottle}$ includes surface loss contributions from the surface of the guide being tested and the gate valve, $\tau_{\beta} = 880.1(1.1)$s is the neutron lifetime \cite{Patrignani2016},  $\tau_{gap}^{-1}$ the loss rate in the gaps between the guide and the shutter, and $\tau_{shutter}^{-1}$ is the loss rate on the shutter.  The leading systematic uncertainty is then the reproducibility of the gap between the guide and the shutter, which has been quoted at 20\%. 

In the measurements presented in \sref{sec:proc} and \sref{sec:ILL} an alternative method of measuring the loss probability was adopted, using two fixed pinhole plates \citep{Makela2014}.  The bottle is created by fixing plates to the ends of the test guide, where each plate has a small hole accounting for $< 10^{-4}$ of the total surface area of the bottle. As shown below, in order for this method to be sensitive to the loss probability of the UCN guide surface, the loss probability through the holes, given by the ratio of the area of the holes to the surface area of the bottle, should be less than or equal to the expected material loss probability, which in this case is $\approx 10^{-4}$.  A gate valve upstream of the bottle is used to control the flow of UCN to the entrance hole.  Downstream of the bottle a detector is closely coupled to the exit hole to monitor the rate of UCN leaving the bottle.  The gate valve is opened for a time $t_{fill}$ equal to several bottle lifetimes to fill the bottle close to the maximum equilibrium density of UCN.   After $t_{fill}$, the upstream gate valve is closed and the rate at which  the density of UCN in bottle decreases is monitored in the downstream detector.  An absorber is mounted to the downstream side of the upstream gate-valve to prevent UCN that exit the bottle through the upstream hole from re-entering the storage volume.   In the ideal case of perfectly lossless coating, no $\beta$-decay, and no gaps, the time dependence of the rate measured by the downstream detector would be similar to charging and discharging a capacitor, with a time constant governed by  (for example, assuming an isotropic UCN velocity distribution, see \cite{UCNBook} p. 91) 
\begin{equation}
\tau_{hole} = 4 V_{guide}/ \langle v \rangle A_{hole},
\label{fig:tauhole}
\end{equation}
where $V_{guide}$ is the volume of the guide, $A_{hole}$ is the surface area of the hole, and $\langle v \rangle$ is the mean UCN velocity. 

Now including the case in which there are other loss mechanisms, the bottle lifetime is determined from
\begin{equation}
	\tau_d^{-1} = \tau_\beta^{-1} + \tau_{bottle}^{-1} + \tau_{gap}^{-1} + \tau_{hole}^{-1},
	\label{eq:bottlet}
\end{equation}	
where $\tau_d$ is the measured drain time of UCN from the bottle after closing the upstream gate-valve.  In these measurements the end plates were coated with the same NiP as the guide and are considered part of the bottle surface area. 

\subsubsection{Extraction of the loss probability from double
pinhole measurements}

For simplicity, frst consider a case where the UCN have a single velocity. Then the UCN loss rate $R_L(t)$ at time $t$ is given by

\begin{equation}
	R_L(t) = \frac{A_{tot}}{4V} \bar{\mu} N(t) v = \frac{\mathrm{d}N(t)}{\mathrm{d}t},
\end{equation}
where $A_{tot}$ is total inner surface area of the bottle, $V$ is the volume of the bottle, and $N(t)$ is the number of the stored UCN at time $t$.  The loss per bounce probability $\bar{\mu}$ can in general depend on $v$ and can also contain contributions from multiple loss mechanisms (including the pinhole contribution).

The rate $R_D(t)$ at which UCN are detected through a pinhole at time t is given by
$R_D(t) = \frac{A_{hole} v }{4 V} N(t),$
where $A_{hole}$ is the area of the pinhole.  $\mathrm{d} N(t) / \mathrm{d}t$ can also be given by
\begin{equation}
	\frac{\mathrm{d}N(t)}{\mathrm{d}t}  = \frac{\mathrm{d} R_{D}(t)}{\mathrm{d}t} \frac{4 V}{A_{hole}v}.
\end{equation}

Then the loss per bounce, the quantity of interest,  is given by
\begin{equation}
	\bar{\mu} = \frac{1}{R_D(t)} \frac{\mathrm{d}R_D(t)}{\mathrm{d}t} \frac{4 V}{v A_{tot}}.
\end{equation}
Thus, if $v$ is known, the detection rate of UCN through the pinhole (and its time dependence) gives the loss probability per bounce.

In general, the UCN stored in a bottle have a velocity distribution.  Therefore, knowing the velocity distribution of the UCN inside the bottle is essential in extracting the loss per bounce information.  The UCN velocity distribution itself can evolve while the UCN are stored. In remainder of this section, we discuss how we obtain the input velocity distribution and our evaluation of the effect of the velocity distribution evolution.

\subsubsection{Velocity distribution from UCN transport Monte Carlo}
\label{sec:velodistr}
A UCN transport Monte Carlo simulation was developed at LANL to assess improvements to the guide system for the source upgrade.  Material interactions in the simulation are parameterized by $\mu_{mc}$, an energy independent loss-per-bounce probability, $f_{mc}$, a loss factor which is related to the loss-per-bounce probability by \eref{eq:loss}, and the probability of having a diffuse reflection, $\eta_{mc}$.  Diffuse scattering occurs in the simulation when a random number sampled uniformly between 0--1 is less than $\eta_{mc}$.  The direction of a UCN exiting a diffuse scatter is sampled from a Lambertian distribution \cite{Clayton2014}.   These parameters can be set independently for each surface in the simulation.  Parameters for the guide system coupling the UCN source to the experimental hall were determined by reconstructing timing distributions measured by A. Saunders {\it et al} \cite{Saunders2013}: $\mu_{mc} = 6.0 \times 10^{-4}$, $f_{mc} = 0$, and $\eta_{mc} = 5.2\%$.  The energy dependent and independent loss probabilities can be used to describe guide sections that are dominated by gaps ($\mu_{mc} \ne 0$,$f_{mc} =0$), sections where the material loss and gaps have similar weight ($\mu_{mc} \ne 0$,$f_{mc} \ne 0$),  or gapless sections ($\mu_{mc} = 0$,$f_{mc} \ne 0$).

\begin{figure}
	\centering
	\includegraphics[width=0.45\textwidth,clip=true,viewport=0 0 550 350]{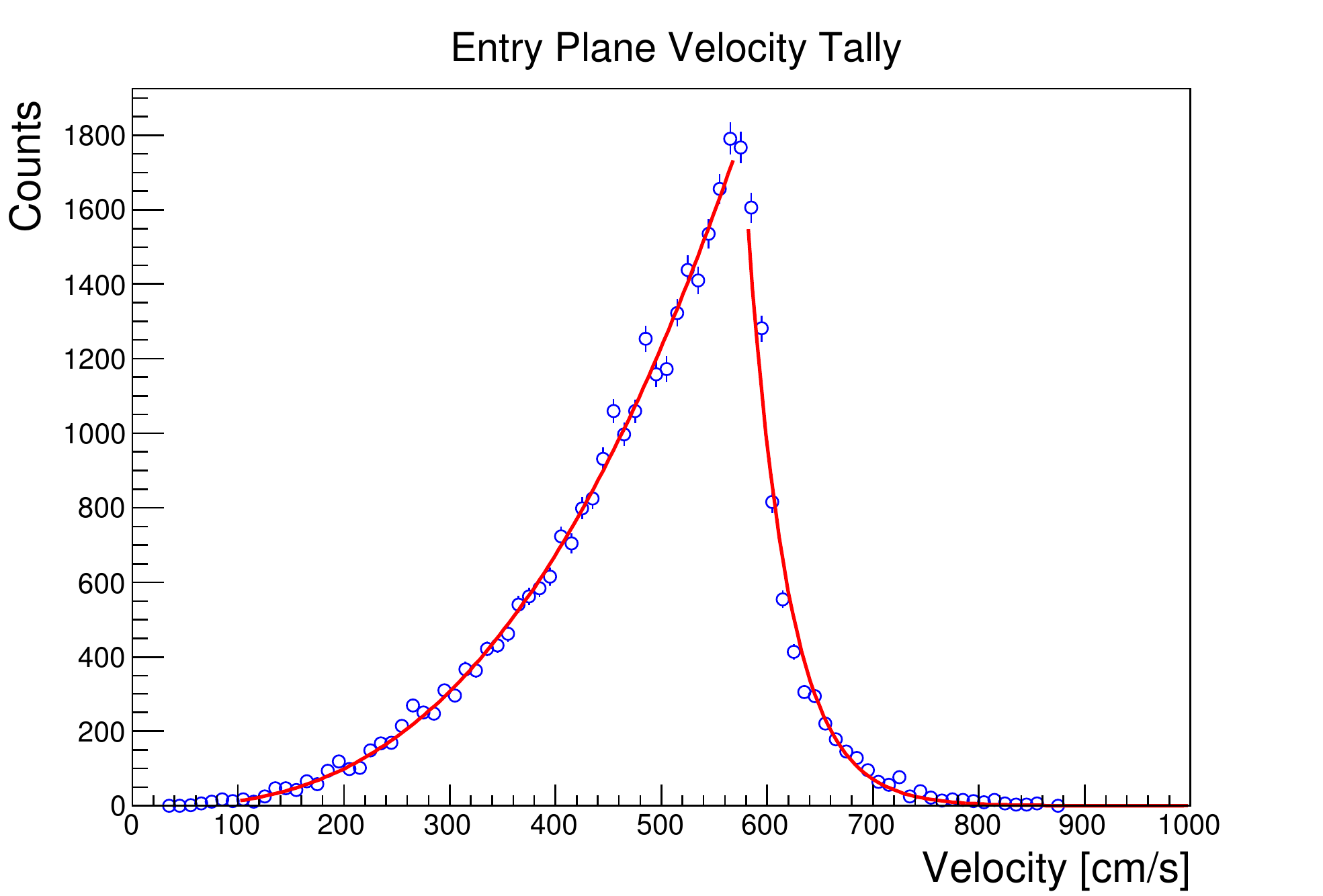}
	\caption{\label{fig:velo} The simulated velocity distribution recorded at the entrance to the test guide for a guide system with an energy independent loss-per-bounce probability of $6\times10^{-4}$ and nonspecularity of 5.2\%. Simulated data are shown as open circles and the curves show the result of a fit to \eref{eq:fitv} (color online).}
\end{figure}

This simulation was used to determine the velocity distribution of UCN entering the test guide for the measurement done at LANL discussed in Sec. \ref{sec:proc}.  UCN with initial kinetic energy up to 400~neV are randomly sampled from a $N(v) \propto v^2$ distribution in the solid deuterium volume of the source and propagated through the guide system to the test guide.  The velocity distribution $P(v)$ of UCN entering the test guide is recorded and fit to a piece-wise function with velocity $v$ above and below $v_c =$5.7~m$/$s
\begin{equation}
	 P(v) =  \left\{
	 	\begin{array}{ll}
	 	a (v-b)^c, & v \leq v_c ,\\
	 	d e^{-v  f}, & v > v_c, 	\\
	 	\end{array}
	 \right.
	 \label{eq:fitv}
\end{equation}
shown in \fref{fig:velo}.  The results of the simulation show that the initial $v^2$ distribution is conditioned during transport and that velocity distribution entering the test apparatus is $\propto v^{2.7}$ with a mean velocity of $\bar{v} =$~4.8(1)~m/s.   These results are consistent with previous analysis of UCN velocity spectrum \cite{Holley2012a}

\subsubsection{Evolution of $\bar{v}$ versus filling time and hole diameter}
\label{sec:vevolve}
Given that the rate at which UCN enter the bottle through a hole is proportional to the velocity, filling the bottle for a finite time will preferentially load faster UCN, in effect heating the spectrum.  This effect is somewhat canceled by the fact that internal loss rate are also proportional to velocity.  A series of transport simulations were performed where the filling time and aperture size were varied to investigate the amount of spectral heating.  

\begin{figure}
	\centering
	\includegraphics[width=0.45\textwidth]{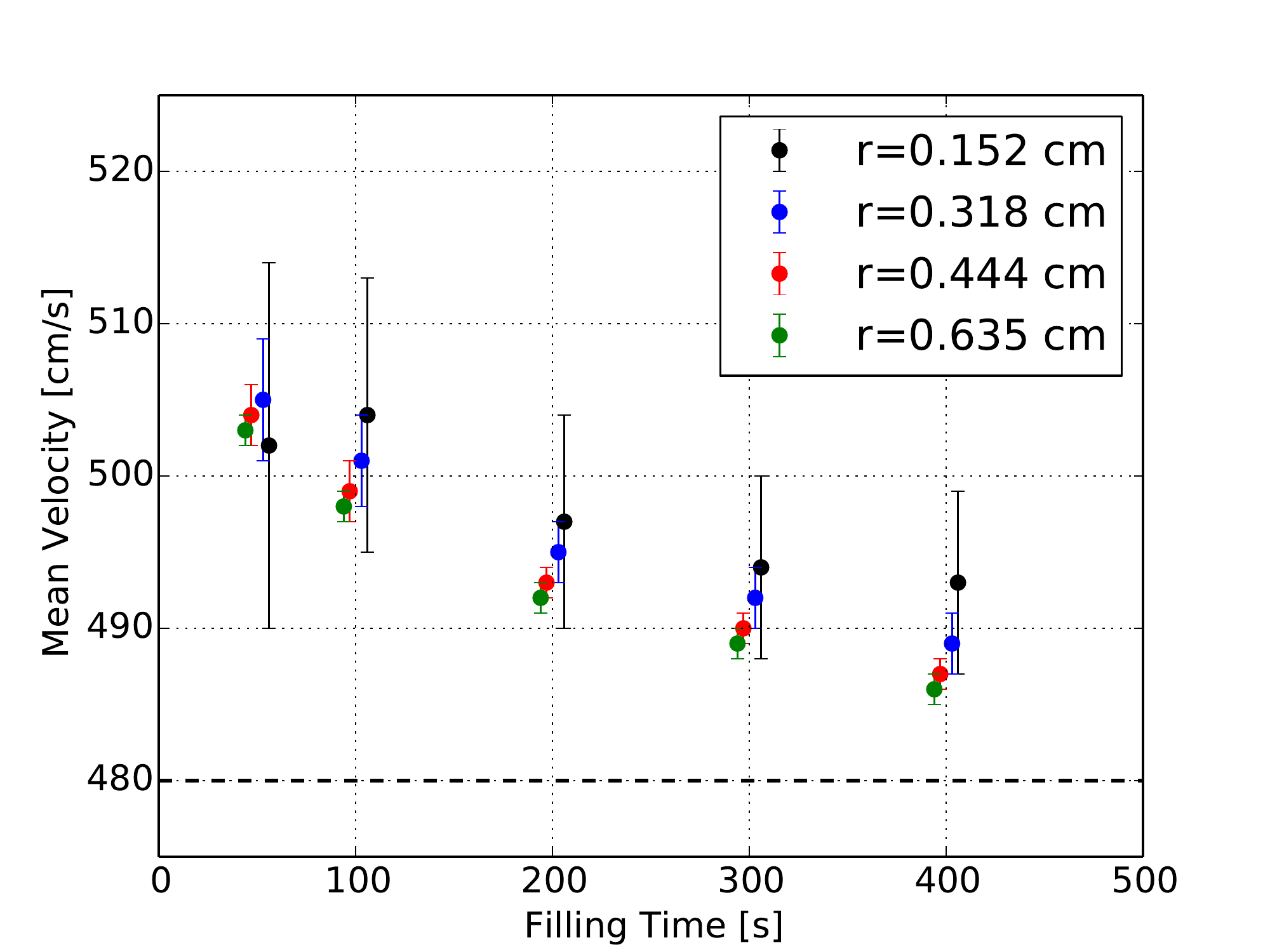}
	\caption{\label{fig:meanv} The mean velocity for UCN in the test bottle as determined by UCN transport Monte Carlo as a function of filling time for hole diameters of 0.304~cm, 0.636~cm, 0.888~cm, and 1.27~cm.  Error bars are statistical (color online). }
\end{figure}

Using the simulation parameters and input velocity spectrum from \sref{sec:velodistr} UCN were simulated filling the bottle for $t_{fill} \in (50,100,200,300,400)$~s through apertures of $r\in(0.152,0.318,0.444,0.635)$~cm. The bottle surfaces had a loss-factor of $f=1.2\times 10^{-4}$, Fermi potential of $212$~neV, and nonspecularity of $\eta=$3\%. A 
cut plane at the exit pinhole was used to record the velocity of UCN that would be detected in the experiment.  UCN tallied at the exit plane are weighted by their probability to surviving until  the time of detection due to $\beta$-decay.  The results of these simulations are shown in \fref{fig:meanv}. As expected, shorter fill time results in a hotter UCN spectra. Also, reducing the radius of the fill hole was observed to increase the mean velocity for a give fill time.  

The measurement presented in the following section use apertures of $r=0.152$~cm and $r=0.318$~cm and a filling time of 300~s. These simulation show that the difference in the internal mean velocity is on the order of a few cm/s or 1-2\% between measurements made with different aperture sizes.  From simulation, the mean velocity of the internal spectrum was found to be 12(3)~cm/s faster than the mean velocity of the input spectrum. 
 
\subsubsection{Time evolution of $\bar{v}$ in a closed bottle}
\begin{figure}
\centering 
 \includegraphics[width=0.45\textwidth]{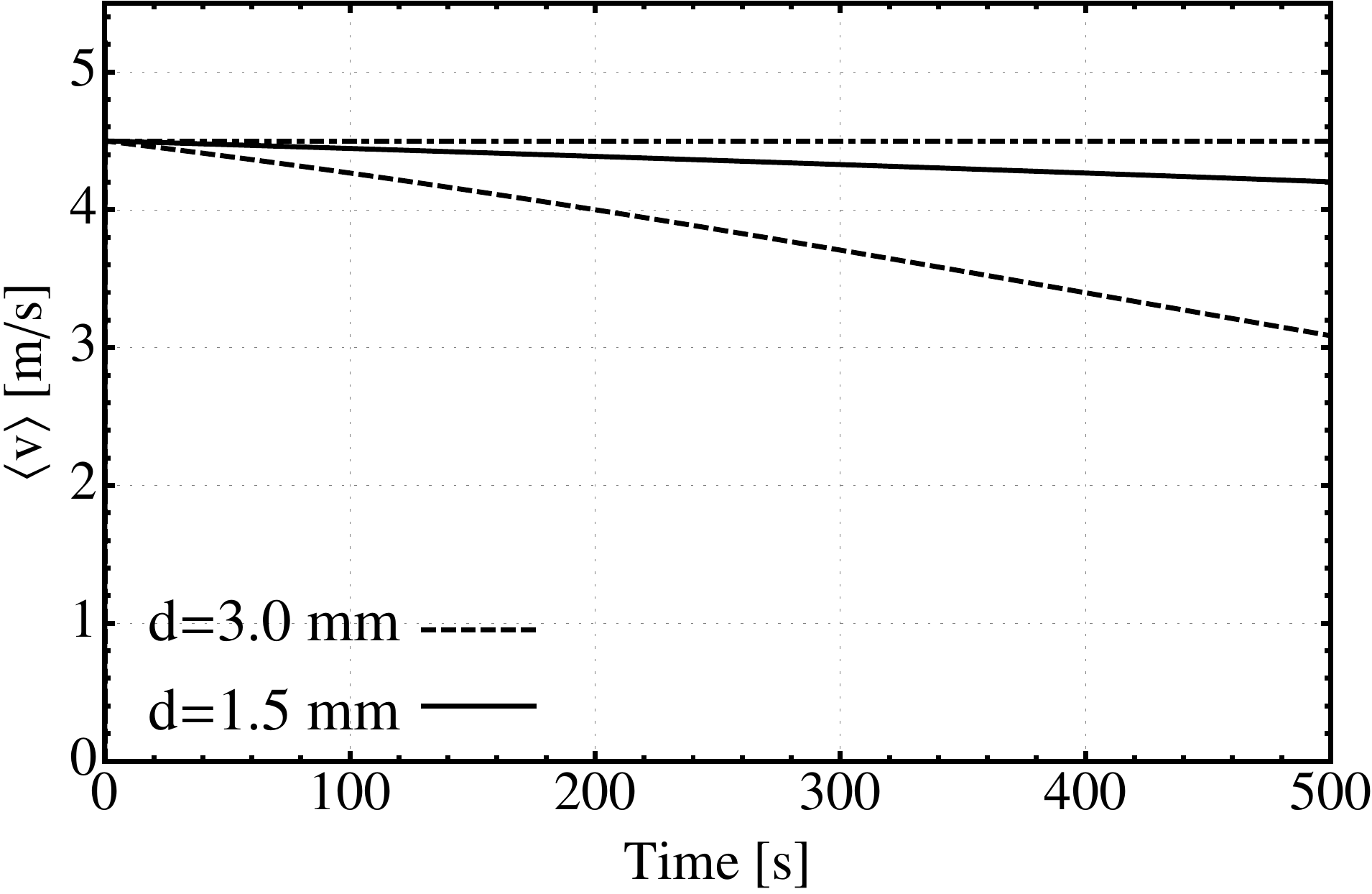}
 \caption{\label{fig:mean_v_t} \textit{Left :} The time dependence of the mean velocity from \eref{eq:tdepv} for $\mathcal{P}(v) \propto v^2$ and a maximum velocity of 6.0~m/s for a 7.5~cm diameter, 100~cm long bottle with drain holes of radius 0.15~cm (solid) and 0.3~cm  (dashed). The dot dashed line represents the mean velocity at t=0. Only losses due to the drain holes are included. }
\end{figure}

As noted by Tang, \textit{et al}\cite{Tang2015} the velocity distribution inside the bottle will evolve with time, cooling the spectrum as the faster UCN are lost through the exit hole or in interactions with the walls.  The time dependence of the mean velocity can be calculated as 
\begin{equation}
	\bar{v}(t) = \frac{\int_0^{v_c} v \mathcal{P}(v)  e^{-t/\tau(v)} \mathrm{d}v} { \int_0^{v_c} \mathcal{P}(v) e^{-t/\tau(v)} \mathrm{d}v},
	\label{eq:meanv}
\end{equation}
where $v_c$ is the cutoff velocity, defined to be twice the maximum trappable velocity of the bottle.  For this analysis the integration was cut off at $v_c\approx 12$~m/s. In most cases relevant to UCN, $\mathcal{P}(v)$ is assumed to be the low energy tail of the Maxwell-Boltzmann distribution. $\mathcal{P}(v)$ is proportional to $v^2$, which leads to an initial mean velocity of $\approx (3/4) v_{max}$.   After substituting  $v^2$ for the velocity distribution, \eref{eq:meanv} can be reduced to
\begin{equation}
\bar{v}(t) = \frac{e^{-b t v_c} \left[ -(b t v_c)^3 - 3 (btv_c)^2 - 6 btv_c - 6 
\right] +6}{b t \left[ e^{-b t v_c} \left\{ -(b t v_c)^2 - 2 b t v_c - 2 \right\} + 2\right]},
\label{eq:tdepv}
\end{equation}
where $1/\tau(v) = bv$, $b = A/4V$ in the case where the hole represents the only loss mechanism in the bottle ($\tau$ given by \fref{fig:tauhole}).  Figure \ref{fig:mean_v_t} shows $\bar{v}(t)$ for $t \in (0,500)$ for a perfect bottle where the drain hole is the only loss.  This analysis shows that treating the drain curve from a pinhole experiment with a static $\bar{v}$ is insufficient to capture the dynamics in the bottle.  Since the velocity distribution is not of the form $\propto v^2$, \eref{eq:meanv} will be numerically integrated in the following analysis.

\subsubsection{Relationship between detected rate and loss rate}

\begin{figure}[ht!]
	\centering
	\includegraphics[width=0.45\textwidth]{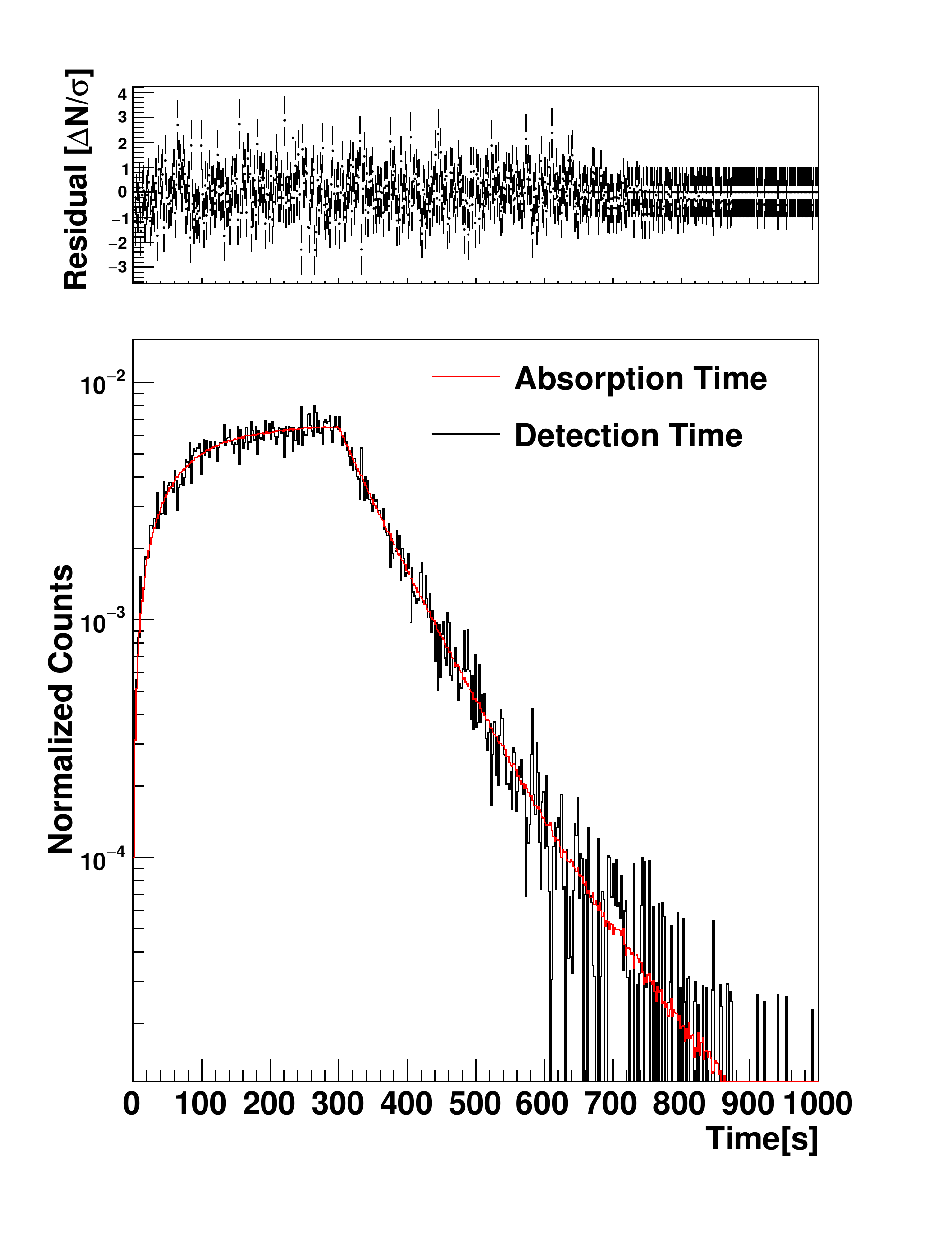}
	\caption{\label{fig:absvdet} The normalized loss rate of UCN on the internal surfaces of the bottle (red line) is plotted against the normalized detection rate of UCN exiting the bottle (black line).   Loss mechanisms accounted for in the simulation include: $\beta$-decay, an energy-independent loss-probability to account for gaps, and an energy-dependent loss-factor to represent interactions with NiP coated surfaces (color online). }
\end{figure}

A possible ambiguity of the double pinhole approach is whether the measured rate of UCN exiting the bottle has a direct relationship with the loss mechanisms internal to the bottle.  This was assessed using a transport Monte Carlo where the gap losses are represented as an energy independent loss-per-bounce probability and an energy dependent loss for the NiP surface scaled by a loss-factor.  The times UCN are lost on the surface of the bottle and the times UCN are detected are tallied. \fref{fig:absvdet} shows that there is a direct correlation between the simulated rate of UCN capturing on a detector downstream of the pinhole and the change of the internal density of UCN. Therefore this method is sufficient for characterizing losses on the material walls of the guide. 

\section{Measurements at LANL}
\label{sec:proc}

\subsection{Experimental Setup}

UCN are produced by cooling spallation neutrons generated by the pulsed 800~MeV proton beam incident on a tungsten target at the LANSCE UCN source \cite{Saunders2013}. A cold moderator made of polyethylene cools a fraction of the spallation neutrons, and these cold neutrons, $E_k \approx 5$~meV, are down-scattered to UCN energies in a solid deuterium (SD$_2$) crystal at $\sim 5$~K.  UCN produced in the SD$_2$ are confined by the Fermi potential of the $^{58}$Ni coated source vacuum vessel.  A series of stainless steel tubes couple to the source volume 1~m above the SD$_2$ volume and guide the UCN through the concrete biological shielding to the experimental area.  This transit conditions the energy spectrum of UCN to the stainless steel Fermi potential, 188~neV.

\begin{figure}
\centering

\includegraphics[viewport= 70 70 340 560,angle=270,clip=true,width=0.45\textwidth]{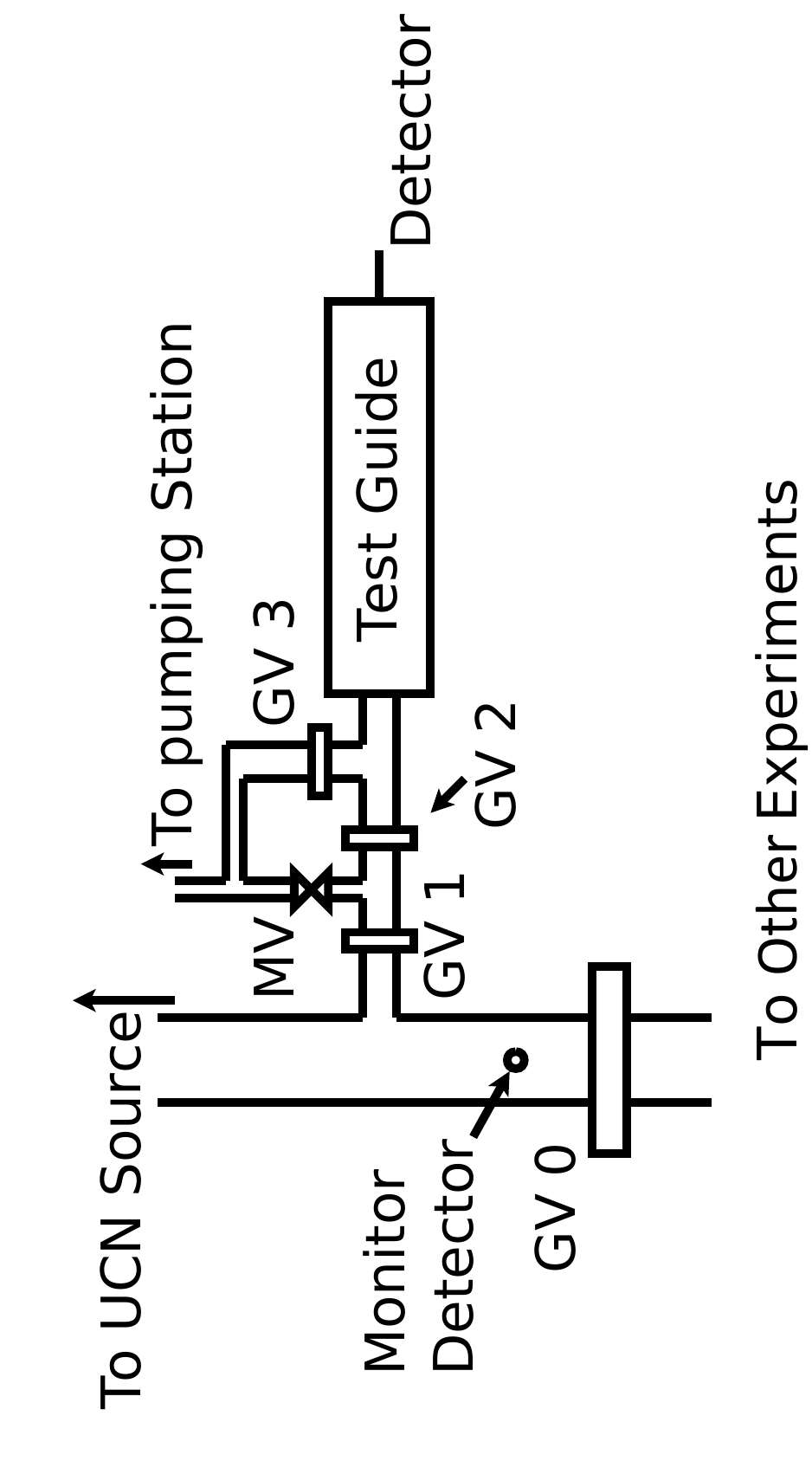}

\caption{\label{fig:gv} Transport and pumping manifold for the guide test apparatus.  Detailed description in the text.}
\end{figure}

The guide test apparatus was coupled to the outlet of the UCN source through a 2.54~cm aperture in the primary UCN guide as shown in \fref{fig:gv}.  In this position the apparatus was mounted upstream of the superconducting magnet which pulls neutrons of one spin state through the vacuum separation foil such that the full output density of the source was accessible.  Experiments mounted in this position run parasitically to the main UCN transport line and some system conditions were controlled by a higher priority experiment downstream.  UCN were loaded into the bottle through the pumping manifold, shown in \fref{fig:gv}.  Valve GV1 separates the holding volume from the source.  Valve GV2 has a thin layer of TPX (polymethylpentene) on the downstream face which, when closed, will absorb and upscatter UCN exiting the test guide through the upstream pinhole.  Valve GV3 allows the test guide to be continually pumped during the counting period.  Valves GV2 and GV3 are controlled by an automated valve control program and are always in opposite states.

Stainless steel pinhole plates (\fref{fig:pinhole1}) were mounted to both ends of the test guide to create a bottle. The interior face of the pinhole plate is flat and the exterior face has a 45$^\circ$ chamfer around the pinhole, which was positioned 2.54 cm off center.  Two sets of pinhole plates were made, one with a 0.635 cm diameter hole and the other with a diameter of 0.318 cm. The up and down stream pinhole plates are mounted to the guide with a 180$^{\circ}$ rotation about the guide axis relative to each other, so that the pinhole was at 6 o'clock on the inlet side and 12 o'clock on the outlet.  All pinhole plates were coated with NiP using the same process as the test guides.

UCN leaving the pinhole on the downstream side of the guide transit through a 2.54~cm diameter outlet port to a $^{10}$B coated ZnS:Ag foil detector \cite{wang2015}. UCN are captured in a 100 nm layer of $^{10}$B coated on the ZnS:Ag substrate.  The neutron capture on the $^{10}$B produces an $\alpha$ and a $^{7}$Li ion and one or both enter the ZnS:Ag substrate creating scintillation light which is measured by a 2.5 cm diameter PMT mounted outside of a vacuum window.  Since the $^{10}$B foil was inside the vacuum system the detection efficiency is relatively independent of UCN energy, unlike the $^{3}$He gas detectors which require a 1 m drop to give UCN sufficient energy to penetrate an entrance window. The UCN density outside the test volume was also monitored with a similar $^{10}$B coated ZnS:Ag detector, which was coupled to the main UCN guide through a 0.159 cm diameter hole.

\subsection{Data Collection}

\begin{figure}[ht!]
\centering
	\includegraphics[width=0.45\textwidth,clip=true]{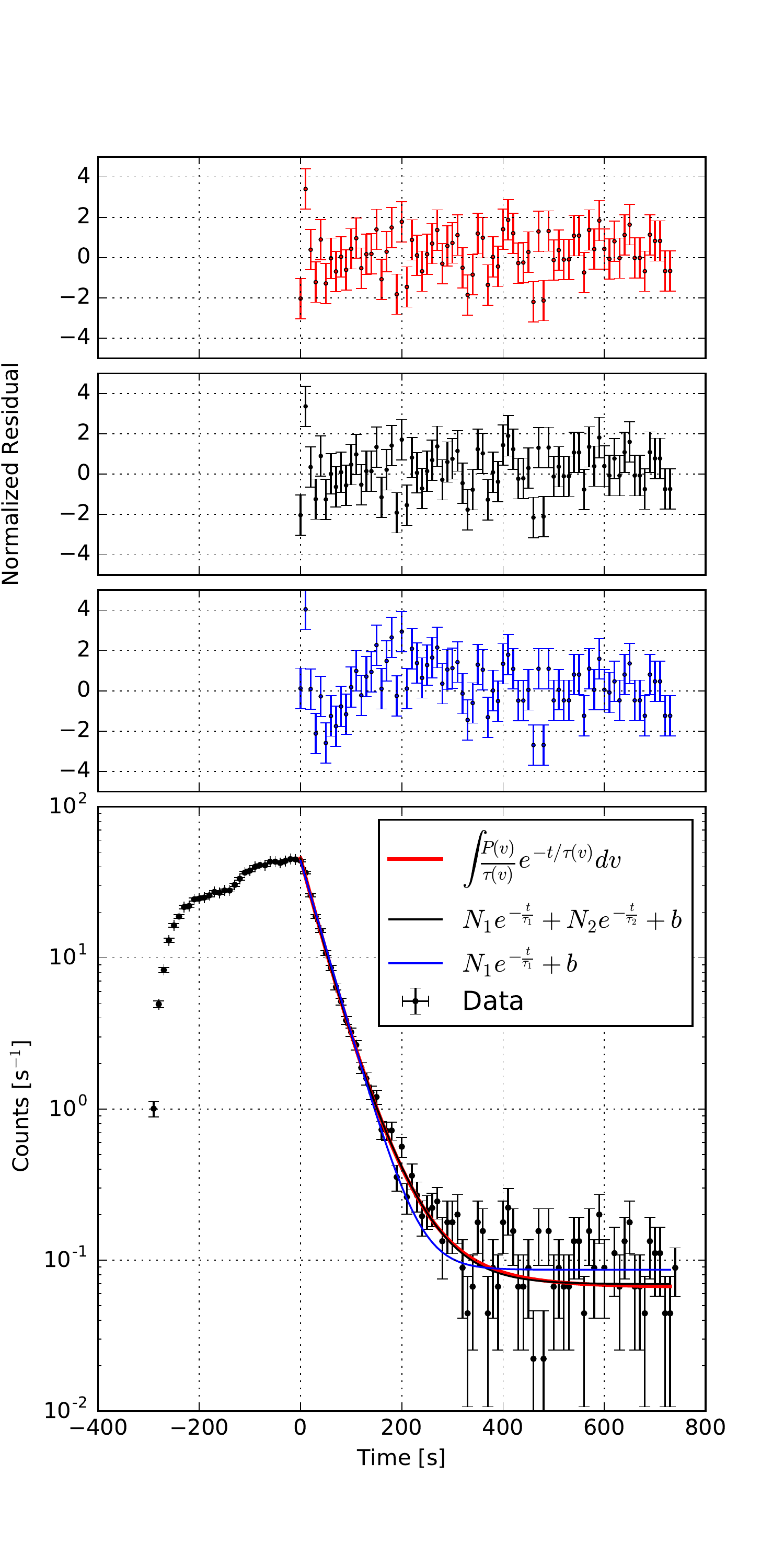}
	\caption{\label{fig:drain} Fill and drain curve from guide 2 with the 0.635~cm diameter pinhole plates.  The increase in count rate between the 0-150~s and 150-300~s period is due to the main gate valve downstream of the experimental setup being closed. Single and double exponential fits to the data are shown as solid blue and black lines respectively. The red line is the result of numerical integration analysis using \eref{eq:rateeq} to fit the data. The upper panels show the normalized residuals of the data to each fit. (color online)}
\end{figure}

The nominal data-taking cycle consisted of filling the bottle for 300~s and then draining for 300~s. These times correspond to several lifetimes for a material bottle with a nominal loss probability of $\mathcal{O}(10^{-4})$ containing UCNs with a mean velocity of $\approx 5$~m/s and mean free path of $\approx 7.3$~cm.  During the fill GV0 was closed (\fref{fig:gv}) after 150~s to accommodate the experiment that was on the main port, which led to an increase in the UCN density outside the test port.  This resulted in an increase in the measured rate in period (150-300~s) relative to the (0-150s) period, shown in \fref{fig:drain} along with various fits discussed in the sections below.  The drain time varied between 270~s and $\sim1000$~s.  Measurements with less than a 600~s fill and drain cycle were analyzed separately and averaged with the longer drain time measurements.  

Signals from the PMT monitoring the $^{10}$B ZnS:Ag foil were amplified with an Ortec 579 Fast Amp and analyzed by an Ortec 551 single channel analyzer.  Timing pulses from the SCA were recorded by a Fast ComTec multichannel scaler.  The measured background rate was typically 0.1~s$^{-1}$ and was determined from a fit for each data set.  The peak signal rate when the UCN reach equilibrium density in guide was 30-40~s$^{-1}$ for the 0.635~cm diameter hole and 4-8~s$^{-1}$ for the 0.318~cm hole.

\subsection{Analysis}
\label{sec:analysis}

The measured rate in the drain detector is fit to a double exponential plus background term of the form
\begin{equation}
 N(t) = N_s e^{-(t-\Delta t) / \tau_s} + N_l e^{-(t-\Delta t)/\tau_l} + B ,
 \label{eq:doublexp}
\end{equation}
where $N_{s,l}$ are the weights of the short and long lifetime ($\tau_s$ and $\tau_l$) components of the UCN population in the bottle, $\Delta t = 300$s is the start of the draining period, and $B$ is the background rate.  The short lifetime term characterizes the UCN which  possess kinetic energy above the bottle potential, typically called super-barrier UCN.  Since it is the perpendicular component of the kinetic energy that determines whether super-barrier UCN will penetrate the surface, they can be trapped so long as they continue to scatter at grazing incidence. Diffuse scattering on the bottle surface can lead to these UCN having closer to normal incidence collisions with the wall and being lost.  The long lifetime term is therefore considered to represent the lifetime of trappable UCN in the bottle and is the related to the loss-per-bounce probability of the material.  The short and long lifetimes determined by fitting the data with \eref{eq:doublexp} are listed in \tref{tab:res}.  

\begin{figure}[hb!]
\centering
\includegraphics[width=0.5\textwidth]{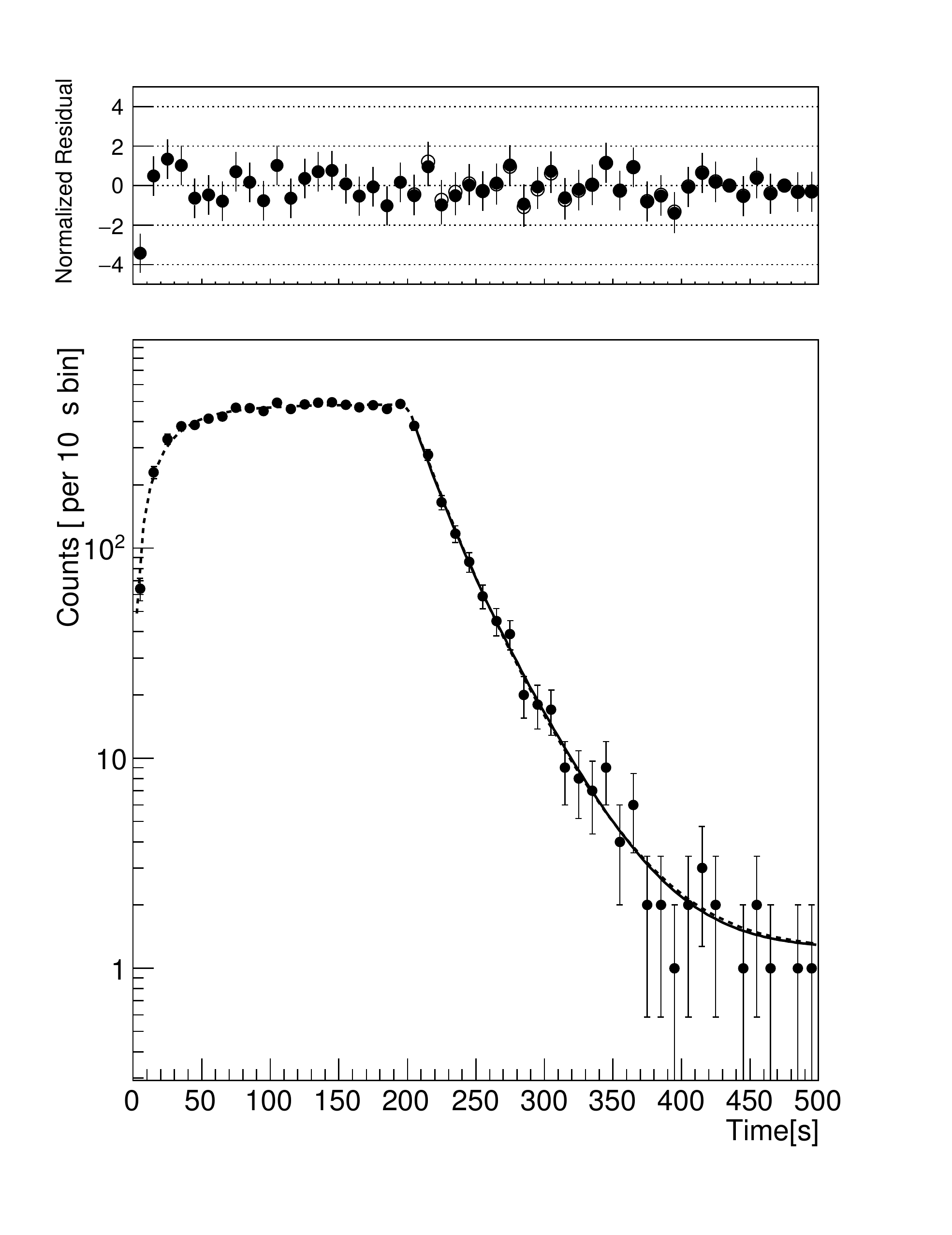}
\caption{\label{fig:drainfitting}  {\it Bottom: }Data for combination of guide 1 and 2 with the 1/4" pinhole plate is shown.  The data is fit using \eref{eq:fitfuncp} for the entire time range (dashed line) and over the draining time (solid line). {\it Top:} The fit residuals normalized by the statistical uncertainty of the bin are shown for a fit of \eref{eq:fitfuncp} over the entire range (solid circles) and over the draining time (open circles).  Error bars are 1$\sigma$.}
\end{figure}

The results of attempting to describe the unloading curve with a single time constant are shown in \fref{fig:drainfitting}. As excepted, the single constant does not represent the data, under predicting the rate at later times.

\begin{table*}
 \centering
 \caption{The short and long lifetimes from a double exponential fit to the draining curve are listed for each guide and pinhole combination, Analysis 1.  The listed loss-per-bounce probability does not correct for the unmeasured effect of the gaps in the system and therefore is the average system loss-per-bounce for the guide and oriface plates. Uncertainties for $\tau_s$, $\tau_l$, $f$, and $l$ are the 1$\sigma$ fit uncertainties.  The resulting loss-factor and gap size from Analysis 2 are shown for indiviudal fits ($f,l$) and the combined fit ($f_c,l$).  Gap lengths that include 0~$\mu$m at 1$\sigma$ are reported as 2$\sigma$ upper bounds. \label{tab:res}}
 \begin{tabular}{cc|rr|r||rr|rr}
 	\hline
 	\hline
 		\T &  &  \multicolumn{3}{c||}{Analysis 1} & \multicolumn{3}{c}{Analysis 2} \\
		\T Guide & Pinhole [cm] & $\tau_s$ [s] & $\tau_l$ [s] & $\bar{\mu}$ [$10^{-4}$] & $f\;[10^{-4}]$ & $l\;[\mu$m] & $f_c$ [10$^{-4}$]  & $l$ [$\mu$m]\\
    \hline
        \T  1 & 0.635 & 19.4(2.0)  & 36.2(0.6) & 2.1(4) &  0.9(1.0) & $\leq 132$   & \multirow{2}{*}{1.1(7)} & 67(63)\\            		
            1 & 0.318 & 22.3(3.2) & 55.8(1.4) & 2.0(3)  & 1.0(1.1)  & $\leq 210$ &  & $\leq 99$\\
      \hline
        \T 2 & 0.635 & 26.2(3.9) & 48.3(2.1) & 1.0(3)   & 1.1(3) & 70(30) &  \multirow{2}{*}{1.5(3)} & 101(28)\\
           2 & 0.318 & 26.4(5.5)	& 73.0(7.6) & 1.4(3)   & 2.7(2) &  $\leq 27$ &  & 31(24) \\
       \hline
      \T   3 & 0.635 & 26.3(1.9) & 67.8(2.5) & 1.1(2)   & 1.8(2) & $\leq 62$   & \multirow{2}{*}{ 1.5(1)} &45(9)\\
           3 & 0.318 & 29.8(4.5) & 85.8(8.7) & 1.4(2)   & 1.1(2) & $\leq 120$ &  & $\leq$ 4\\
         \hline
    \T   1+2 & 0.635 & 33.3(2.0)  & 58.8(1.6) & 1.5(2)  & 2.3(2) & $\leq 120$  & \multirow{2}{*}{1.2(1)}&18(7)\\
         1+2 & 0.318 & 27.4(11.2) & 74.7(4.0) & 1.6(2)  & 1.9(4) & $\leq 60$ & & $\leq$ 9\\
    \hline
    \hline
 \end{tabular}
\end{table*}

\subsubsection{Analysis 1: Double exponential fitting}
\label{sec:dblefit}
The bottle lifetime $\tau_{bottle}$ is related to the measured drain time, $\tau_d$ by 
\begin{equation}
 \tau_{d}^{-1} = \tau_{bottle}^{-1} +  \tau_{\beta}^{-1} + \tau_{hole}^{-1} +\tau_{gaps}^{-1},
 \label{eq:lifetime}
\end{equation}
where $\tau_{\beta}$ is the neutron lifetime, $\tau_{hole}$ is the drain time of the pinholes, and $\tau_{gaps}^{-1}$ is the loss rate of UCN in gaps between the end of the test bottle and pinhole plates or guide to guide mating.  The drain time of UCN out of a hole is $\tau_{hole} = 4 V / (\langle v \rangle A_h)$, where $V$ is the volume of the bottle, $A_h$ is the area of the hole, and $\langle v \rangle$ is the average velocity of the UCN in the bottle.   The rate of loss due to gaps between the guide end and pinhole plate is estimated from
\begin{equation}
 \tau^{-1}_{gap} = \epsilon \frac{\langle v \rangle A_{gap} }{4 V} = \epsilon \frac{\langle v \rangle  \langle d \rangle  }{   r l },
\end{equation}
where $\epsilon$ is probability that UCN that find the gaps to be lost, $\langle d \rangle$ is the average gap size, $r$ is the radius of the guide, and $l$ is the length of the guide.  If a gap is sufficiently diffuse the loss probability should approach 50\%, however for this analysis a conservative approach is taken and $\epsilon$ is set to 1 and $\langle d \rangle$ is the effective gap size.

The bottle lifetime can be written as
\begin{equation}
 \tau^{-1}_{bottle}= \frac{\langle v \rangle A_T \bar{\mu}}{4 V},
 \label{eq:taubottle}
\end{equation}
where $A_T$ is the surface area of the bottle and $\bar{\mu}$ is the loss probability averaged over the ensemble of UCN.  \eref{eq:lifetime} can be rewritten as
\begin{equation}
\tau_{d}^{-1} = \tau_\beta^{-1} + \frac{\langle v\rangle }{4 V} (A_T \bar{\mu} + A_h).
\label{eq:drain2}
\end{equation}
Here $\bar{\mu}$ is assumed to be an averaged system loss rate incorporating loss on the bottle walls as well as loss due to gaps between the tube and the end plates.  The Monte Carlo generated velocity distribution from Section \ref{sec:velodistr} is used to determine the mean velocity $\langle v \rangle=4.8$~m/s used in this analysis. The loss probability for the bottle system $\bar{\mu}$ can be determined using the measured $\tau_l$ and are listed in \tref{tab:res}.  The error-weighted mean loss-probability from the measurements performed on the four guides is $\bar{\mu} = 1.3(1)\times 10^{-4}$.

If the difference in $\tau_l$ measured in guide 3 and the guide 1+2 system is attributed to the additional two gaps created by the guide coupling plate the size of the internal gaps can be estimated.  The total additional gap is $<76 \mu$m, assuming 100\% loss in the gap.  

\subsubsection{Analysis 2: Numerical integration fitting}
\label{sec:numint}
An alternative analysis method is to fit the data to
\begin{equation}
	\frac{\mathrm{d}N(t)}{\mathrm{dt}} = N_0 \int_0^{v_{cut}}  \frac{P(v)}{\tau(v)} e^{-t/\tau(v)} \mathrm{d}v +B,
	\label{eq:rateeq}
\end{equation}
which incorporates the full velocity dependence of the \eref{eq:lifetime} and the input velocity distribution $P(v)$.  Instead of fitting the data to multiple time constants the fit parameters for \eref{eq:rateeq} are the overall normalization $N_0$, the loss-factor $f$, the total gap length $l$, and the background rate $B$.  Because of the complicated dependence on energy in $\mu(E)$ \eref{eq:rateeq} is numerically integrated up to $v_{cut}=$12~m/s and evaluated at each time step. 

Two approaches were used in this analysis.  The results of each approach is summarized in \tref{tab:res}.  In the first approach \eref{eq:rateeq} was fit to the data with $N_0$, $f$, $l$, and $B$ as free parameters, where all parameters were constrained to be positive.  In this approach the loss-factor and gap size were found to be strongly correlated, but in some cases the fit algorithm failed to find a non-zero gap size. From averaging the results from the five data-sets with non-zero gap sizes, the mean loss-factor and gap length are $f=1.4(1)\times 10^{-4}$ and $l = 50(22)$~$\mu$m.  In the second approach, the drain curves obtained for the two radii pinholes were simultaneously fit for each guide, constraining the loss-factor $f_c$ to be a shared parameter and $N_0$, $l$, and $B$ as free parameters. This approach yielded an average loss-factor for the NiP coating of $\langle f_c \rangle = 1.36(7)\times 10^{-4}$.

\section{Measurements at the ILL}
\label{sec:ILL}

\subsection{Experimental Setup and Data Collection}

The guide test apparatus was mounted on the TEST port of the UCN turbine PF2 at the ILL Grenoble, France \cite{Steyerl1986}.  Measurements performed at PF2 include the characterization of the UCN velocity spectrum, the relative transmission of electropolished and nonelectropolished NiP coated guides, and the double pinhole bottling lifetimes. All measurements were performed downstream of a 100~$\mu$m Al foil which separates the turbine vacuum from the experimental volume, blocking neutrons with kinetic energy below the aluminum Fermi potential $V_F^{Al} = 54$~neV.
 
Ultracold neutrons were detected with the same $^{10}$B ZnS:Ag foil detectors used in the LANL measurements.  Signals from the photo-multiplier tube monitoring the foil were amplified with an Ortec 474 timing filter amp and recorded with CAEN DT5724 100 MHz, 14-bit analog to digital converter digitizer. 

\subsubsection{Velocity Spectrum Measurements}

\begin{figure}[t]
\centering
\includegraphics[width=0.5\textwidth,clip=true,viewport=70 500 550 720]{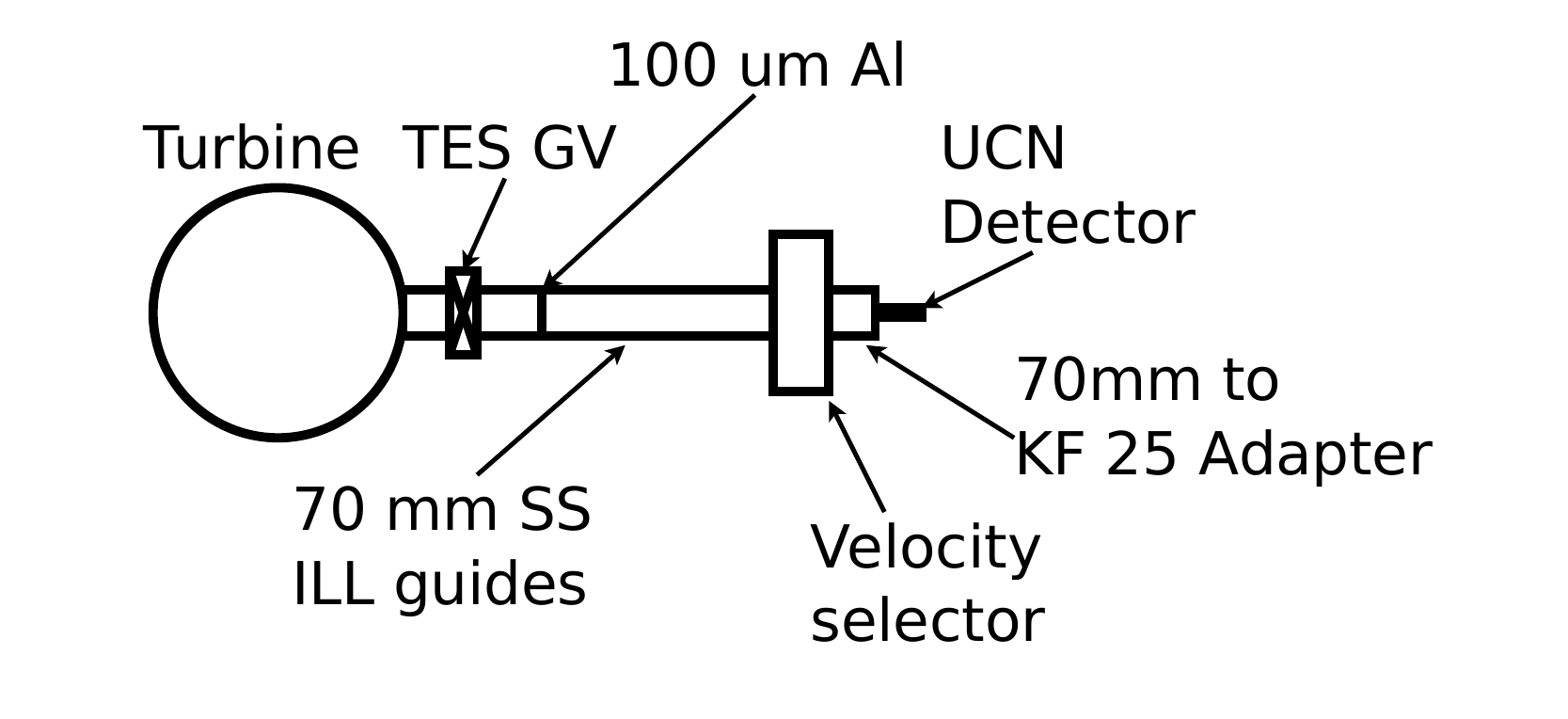}
\caption{\label{fig:velo_selec} The velocity selector was connected to the turbine by a 1 m long 70 mm OD stainless guide after the coupler housing the 100~$\mu$m thick Al safety foil. On the downstream side a 22 cm long 70 mm OD to KF-25 adapter was used to attach the UCN detector. }
\end{figure}

\begin{figure}[t]
\centering
	\includegraphics[width=0.5\textwidth]{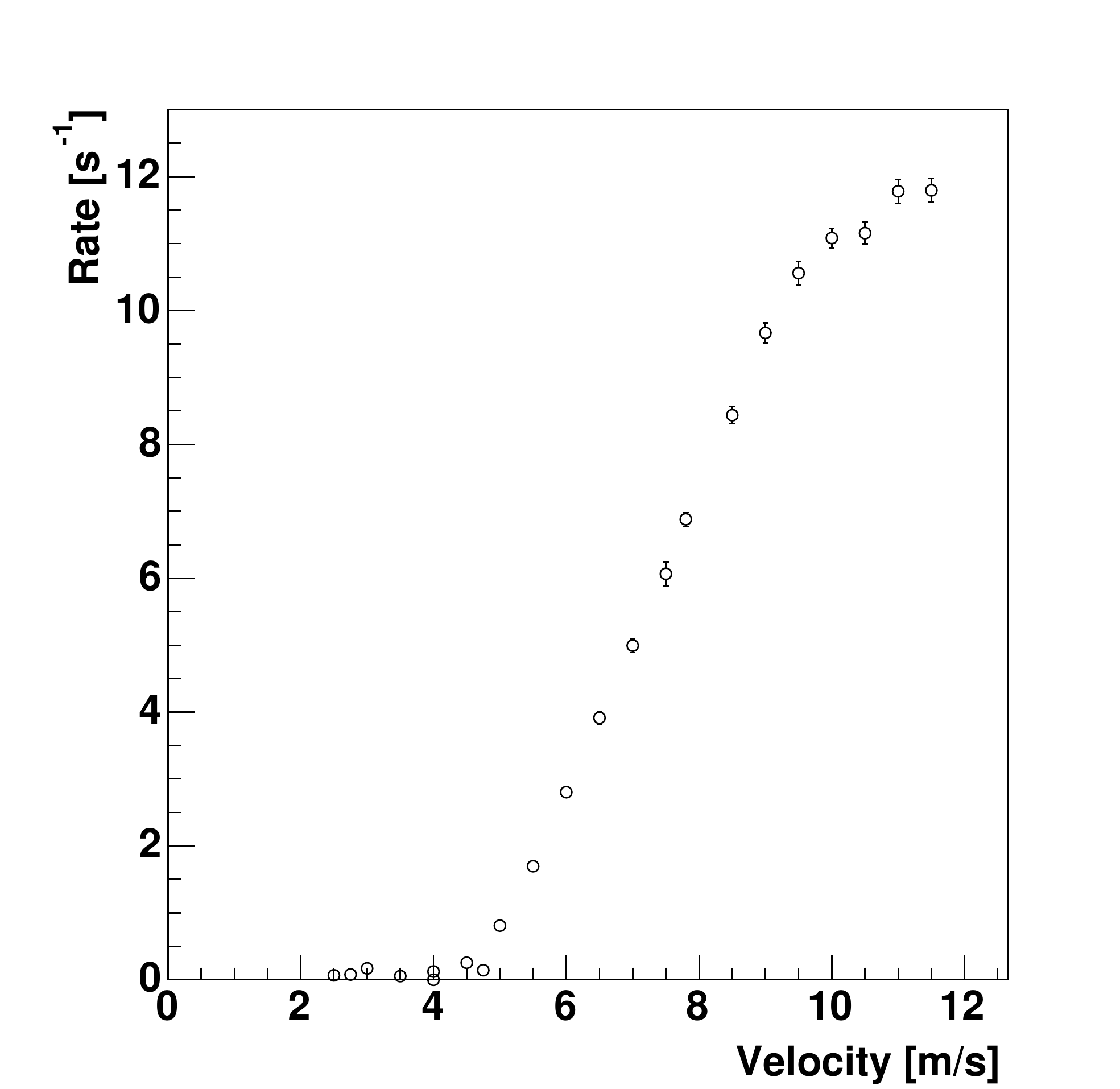}
	\caption{\label{fig:velospec} Velocity spectrum measured at the ILL TEST	 port using the TUM velocity selector.}
\end{figure}

The velocity spectrum of UCN exiting the test port was measured using a rotating disc velocity selector \cite{Schroff2016}.  The velocity selector has seven titanium discs with 26 evenly spaced 19$^{\circ}$ slits. The slits cover 50\% of the surface area of each disc. The discs are positioned such that, when spinning at frequency $f$, UCN with velocity $v = f/5$ are transmitted.   A diagram of how the velocity selector was  attached to the beam line is shown in \fref{fig:velo_selec}.  The velocity spectrum was measured between 1-11.5~m/s in 0.5 m/s steps and the results are shown in \fref{fig:velospec}.  The results of this measurement confirm that UCN with velocities below $\approx$ 4.5~m/s are not transmitted through the aluminum safety foil.  NiP coated guides having a Fermi potential of $V_F = 213$~neV can trap UCN with $v_{max} < 6.3$~m/s and therefore the majority of the neutrons at this position on the turbine are non-trappable.  The measured velocity distribution was used as an input the Monte Carlo simulations of the test geometry.

\subsubsection{Relative Transmission Measurements}

\begin{figure}[ht!]
\centering
\includegraphics[height=0.45\textwidth,angle=270,clip=true,viewport=100 100 350 680]{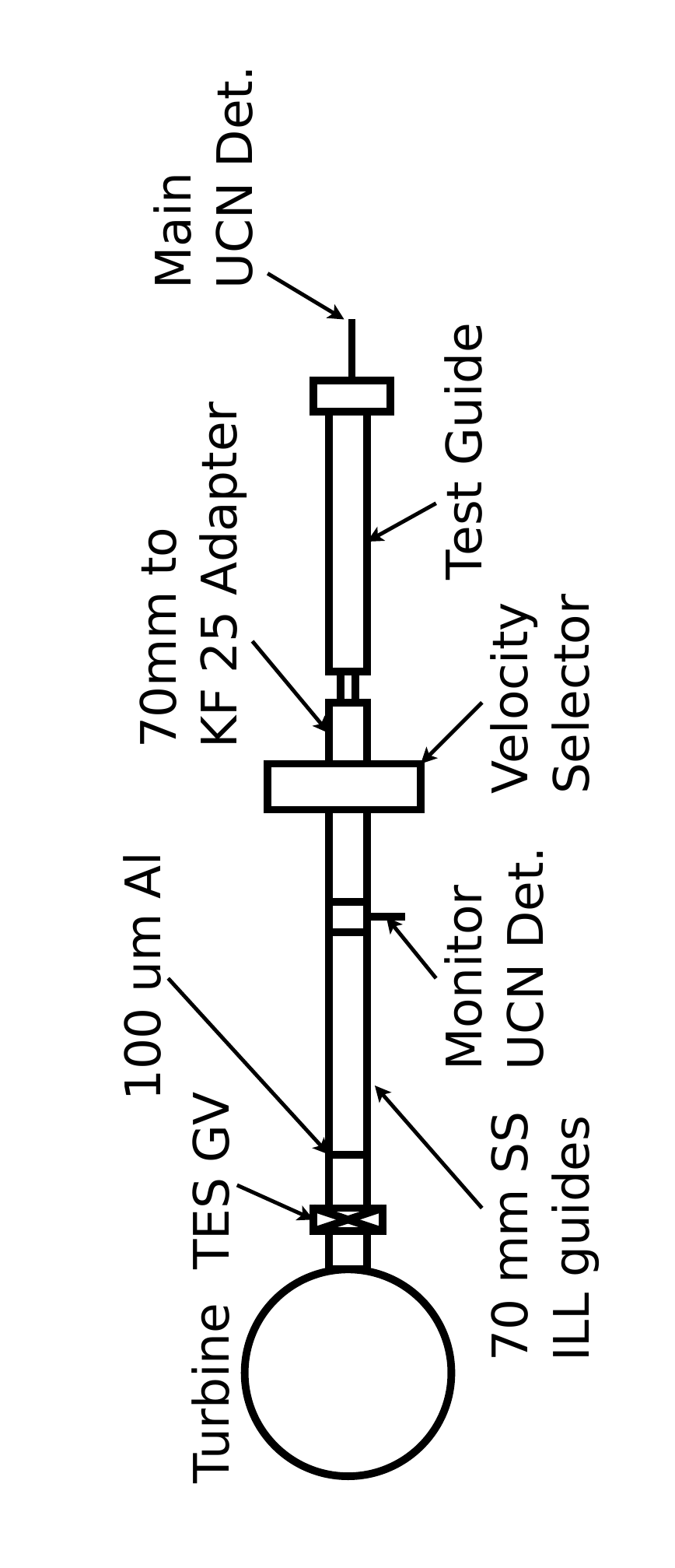}
\caption{\label{fig:transmission} Guide configuration for the measurement of transmission through NiP coated guides. }a
\end{figure}

The relative transmission through an electropolished and non-electro\-polished 140 cm length of guide was measured for UCN with $v=5.5$~m/s.  The test configuration is shown in \fref{fig:transmission}.  Upstream of the velocity selector the incident flux of UCN is monitored through a 64~mm diameter hole in the wall the guide coupler.  A 70~cm length of guide (guide 1) was mounted downstream of the velocity selector followed by a 140 cm length of guide.  Both 140 cm guides were coated with 50 $\mu$m of NiP, however guide 4 was electropolished prior to coating. The rate at the end of the guide was measured using a $^{10}$B ZnS:Ag detector mount on a 70 mm to KF25 adapter.  A thin sheet of TPX with a 2.54 cm diameter hole was fixed to the interior of the adapter to prevent bottling, preventing multiple bounces which distort the result.

The rate measured in the main UCN detector was normalized to the rate in the upstream monitor and the background was measured by closing the turbine shutter.  The background-subtracted normalized rates were $2.4(3) \times 10^{-3}$ and $1.4(3)\times10^{-3}$ for the electropolished and non-electro\-polished guides respectively, corresponding to relative transmission of 55(14)\%.  The statistical uncertainty baseline measurement was too large to determine an absolute transmission, however the relative measurement of guides 3 and 4 confirms the vendor's observation that the NiP coating enhances inherent surface roughness, degrading the transmission.  It has been noted that coating process used by other vendors does not amplify the roughness of the substrate.

\subsubsection{Bottle lifetime data collection}

\begin{figure}
\centering
 \includegraphics[height=0.45\textwidth,viewport=100 100 350 600,clip=true,angle=270]{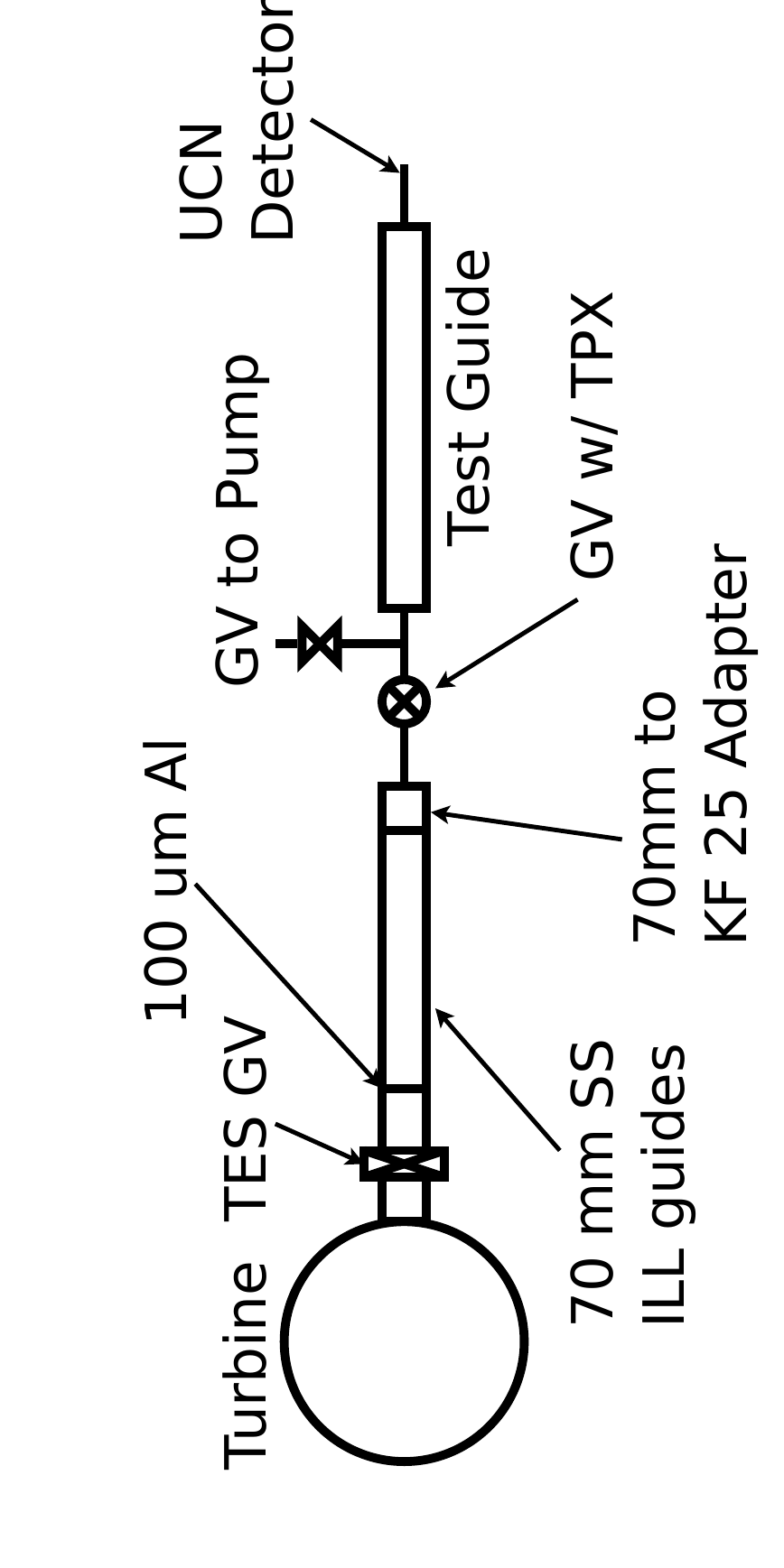}
 \caption{\label{fig:bottlesetup} Bottling measurements were performed using a gate valve manifold system that couples to the feed guide and the pinhole plate with KF-25 connections.  The gate valve "GV /w TPX" is opened to filled the test bottle and when closed introduces a TPX foil to the upstream side of the test guide, removing any neutrons escaping on that side.  The gate valve "GV to Pump" allows continuous pumping of the test guide and its state is always opposite of the inline gate valve. }
\end{figure}

The bottling lifetime was measured by filling the test guide through small aperture for a time $\Delta t = 200$~s and monitoring the count rate through a similar aperture on the downstream side of the test guide.  The aperture plates were the same as those used \sref{sec:proc}. After filling, the gate valve was closed  and the rate of UCN exiting the downstream aperture was monitored for 300~s.  A time tag in the data, generated when the gate valve opens, allows the beginning of the filling period to be aligned in analysis. The 70 mm OD ILL guide from the turbine was connected to gate valve manifold system, via 70 mm to KF-25 reducer, which allowed the flow of neutrons to be stopped and introduced a TPX foil on the upstream side of the test guide to remove UCN escaping from that side of the test guide.  This prevented reloading of the test guide during the draining period.  Additionally, there was a second gate valve, not in the path of neutrons, which allowed continuous pumping on the bottle.  Typically, the rate of UCN detected downstream of test guide was 10 s$^{-1}$ when the bottle had reached its maximum density. Therefore, to achieve sufficient statistical precision each guide was measured for approximately eight hours or fifty fill and drain cycles.

\subsection{Analysis}

\subsubsection{Double Exponential Analysis}

The rate that UCN drain from the test bottle can be characterized to first order as
\begin{equation}
 N(t) = \left\{ 
 \begin{array}{ll}
		\sum_{i=l,s} \left( N_i(1-e^{-t/\tau_i}) \right) + B, & t < t_{fill}\\ 
        \sum_{i=l,s} \left( N_i e^{-(t-t_{fill})/\tau_i} \right) + B ,& t > t_{fill} \\
 \end{array}
 		  \right.
 		  \label{eq:fitfuncp}
\end{equation}
where $t_{fill} = 200$ s, $N_i$ is normalization parameter, $\tau_i$ is the time constant of the system, the indices $l$ and $s$ denote the long and short components of the time constant, $B= n_c R_{bkg}$ is the fixed background rate, where $n_c$ is the number of fill and drain cycles, and the measured background rate $R_{bkg}=0.006(1)$~s$^{-1}$ for the 70 cm guides and $R_{bkg} = 0.0035(4)$~s$^{-1}$ for the 140 cm guides.  In this analysis the filling and draining time constants were assumed to be the same, $\tau_{F} = \tau_{D}$. The short and long time constants are interpreted in the same way as in \sref{sec:dblefit}. The results of this analysis are summarized in \tref{tab:illres} and the mean loss-probability is $\bar{\mu} = 1.6(2)\times 10^{-4}$. 

\begin{table*}[ht]
	\centering
	\caption{\label{tab:illres} The results of fitting \eref{eq:fitfuncp} to the fill and drain timing data and the loss per bounce probability are summarized below.  Fit results which include 0 at 1$\sigma$ are reported as 2$\sigma$ upper bounds.}
	\begin{tabular}{c|rr|r||rr||r}
	\hline
	\hline
	\T \B Guide & $\tau_s$ [s] & $\tau_l$ [s] & $\bar{\mu} [10^{-4}]$   & $f\;[10^{-4}]$ & $l\; [\mu m]$ & $\bar{f} \; [10^{-4}]$ \\
	\hline
	\T	1   & 19.5(4.3) & 45.3(20.1) & 2.0(1.2) & $\leq 2.0$& 178(23)     & 1.7(1) \\
		2   & 26.4(8.3) & 59.4(12.3) & 1.4(0.4) & $\leq 2.1$& 92.7(8.1)   & 1.0(1) \\
		1+2 & 28.3(9.1) & 57.5(18.2) & 1.7(0.7) & $\leq 2.3$& $\leq 460$  & 1.2(1) \\	
	\B	3   & 19.1(5.5) & 62.2(8.9)  & 1.6(0.3) & 1.2(1)    & $\leq 140$  & 1.2(1) \\
	\hline
	\hline	
	\end{tabular}
\end{table*}

\subsubsection{Numerical Integration Analysis}
The same analysis method used in \sref{sec:numint} were applied to the ILL data using the measured velocity spectrum.  Results for the fit parameters $f$ and $l$ for the two parameter fit and $\bar{f}$ for the one parameter fit are summarized in \tref{tab:illres}.  The entire data-set failed to converge on values for the loss-factor and gap length in the two parameter fit.  To mitigate the instability of the two parameter fit, the gap length was fixed to zero and an effective loss-factor was used to fit the data.  In this method the effect loss-factor is $\bar{f} = 1.3(1) \times 10^{-4}$.

\section{Discussion}
\label{sec:disc} 

The results presented in \tref{tab:res} and \tref{tab:illres} show that the system loss-probability for UCN interacting on stainless steel guides coated with 50~$\mu$m of nickel phosphorus and coupled with the design described in \sref{sec:guide} is $\bar{\mu} = 1.4(1)\times 10^{-4}$.  This new guide coupling flange has shown to reliably and repeatably join guides with minimal gaps improving the bottle lifetime over previous designs using ConFlat flanges.  Because the gaps can be reliably small this opens the possibility of using multiple sections of short guide to cover long distances which allows for the guides to be coated using methods that typically have length limitations (such as NiP and NiMo). 

Analysis of ILL and LANL data-sets based on the method discussed in \sref{sec:dblefit} produced consistent values for $\bar{\mu}$ after accounting for the different velocity distributions at each UCN source.  Using a numerical approach to the analysis, this data resulted in a consistent loss-factor, $f_c = 1.3(1) \times 10^{-4}$ when a combined fit to the small and large pinhole drain curves was used.  Nickel phosphorus guide coating will provide a significant improvement in UCN density over stainless steel for experiments which fill a storage volume such as nEDM and neutron lifetime measurements.  Results from the ILL indicate that the transmission through the electropolished nickel phosphorus coated guides are better than 85\% at the 95\% confidence level, but another measurement is required to determine the absolute transmission.
 
The pinhole bottling method of characterizing surfaces provides a complementary approach to the gate valve bottling method with advantages and disadvantages.  In both methods knowledge of the initial velocity spectrum of UCN in the bottle and how that spectrum evolves with time is required to extract relevant surface parameters.  One advantage of this method is that it requires no moving parts removing any uncertainties that arise due to mechanical reproducibility. However, loading and draining the bottle through the static pinhole increases the time required to make a measurement and as guides improve (loss-factor decreases) the fill and drain cycle will take longer further increasing the time required.  

Combining this work with the results of our earlier work \cite{Tang2015} indicates that nickel phosphorus is a commercially available, low-depolarizing, low-loss coating for ultracold neutron transport guides. These qualities make NiP coatings ideal for large UCN transport systems and have already been implemented in the upgrade of the LANL source. The electroless coating process allow parts with complex geometries to be uniformly coated.

\section{Conclusions}

We have measured the storage time of UCN in a nickel phosphorus coated bottle at two facilities with different neutron velocity distributions.  The results of these measurements are used to determine an effective loss-per-bounce probability in the bottle of $\bar{\mu}<1.6 \times 10^{-4}$, which fulfills the transport requirements for upgrade of the LANL UCN source. The results of this measurement demonstrate that the pinhole method is a reliable technique to characterize UCN surface properties. The Fermi potential of the NiP coating, $V_F = 213(5)$~neV, was measured at the ASTERIX time-of-flight spectrometer.  Transmission measurements determined that although the guide is electropolish in production a second electro-polishing after welding the flanges on is required to maintain a specular surface.

\section*{Acknowledgements} 
This work was supported by the Los Alamos National Laboratory LDRD office (DR-20140015DR), the Department of Energy (DE-FG02-97ER41042,), and the National Science Foundation (1307426) and is covered by LA-UR-17-20247.  These measurements would not have been possible without the effort of the LANSCE accelerator staff and the ILL reactor crew.

\bibliography{nimo_guide}
\bibliographystyle{apsrev_nourl}
\biboptions{sort&compress}

\end{document}